\documentclass{aa}  
\usepackage{graphicx}
\usepackage{amssymb}
\usepackage{latexsym}
\usepackage{latexsym}
\usepackage{graphics}
\usepackage{txfonts}
\usepackage{natbib}

\def\ha{H$\alpha$}
\def\hb{H$\beta$}
\def\hy{H$\gamma$}

\def\niib{[NII]$\lambda$6584}

\def\nii{[NII]$\lambda$6548,6584}

\def\oiiib{[OIII]$\lambda$5007}
\def\oiii{[OIII]$\lambda$4958,5007}

\def\siiab{[SII]$\lambda\lambda$6716,30}

\def\heii{HeII$\lambda$4686}
\newcommand{\MOKA}{\ensuremath{\mathrm{MOKA^{3D}}}}

\usepackage{txfonts}
\begin{document} 

   \title{Bubbles and outflows: The novel JWST/NIRSpec view of the z=1.59 obscured quasar XID2028} %\thanks{This work is based on observations made at...}

   \author{G. Cresci\inst{1} \and
          G. Tozzi\inst{2,1} \and
          M. Perna\inst{3} \and
          M. Brusa\inst{4,5} \and
          C. Marconcini\inst{2,1} \and
          A. Marconi\inst{2,1} \and
          S. Carniani\inst{6} \and
          M. Brienza\inst{5,4}  \and
          M. Giroletti\inst{7} \and
          F. Belfiore\inst{1} \and
          M. Ginolfi\inst{2,1} \and
          F. Mannucci\inst{1} \and
          L. Ulivi\inst{2,1} \and
          J. Scholtz\inst{8,9} \and
          G. Venturi\inst{10,1} \and
          S. Arribas\inst{3} \and
          H. {\"U}bler\inst{8,9} \and
          F. D'Eugenio\inst{8,9} \and
          M. Mingozzi\inst{11} \and
          B. Balmaverde\inst{12} \and
          A. Capetti\inst{12} \and
          E. Parlanti\inst{6} \and
          T. Zana\inst{6}   
          %\thanks{Just to show the usage of the elements in the author field}
          }

    \institute{INAF - Osservatorio Astrofisco di Arcetri, largo E. Fermi 5, 50127 Firenze, Italy \\ \email{giovanni.cresci@inaf.it}  
        \and
        Universit\`a degli Studi di Firenze, Dipartimento di Fisica e Astronomia, via G. Sansone 1, 50019 Sesto F.no, Firenze, Italy
        \and
        Centro de Astrobiología (CAB, CSIC–INTA), Departamento de Astrofísica, Cra. de Ajalvir Km. 4, 28850 – Torrej\'on de Ardoz, Madrid, Spain
        \and
        Dipartimento di Fisica e Astronomia, Alma Mater Studiorum, Universit\'a di Bologna, via Gobetti 93/2, 40129 Bologna, Italy
        \and
        INAF - Osservatorio di Astrofisica e Scienza dello Spazio di Bologna, via Gobetti 93/3, 40129 Bologna, Italy
        \and
        Scuola Normale Superiore, Piazza dei Cavalieri 7, I–56126, Pisa (PI), Italy
        \and
    INAF - Istituto di Radioastronomia, via Gobetti 101, 40129 Bologna, Italy
    \and
        Cavendish Laboratory, University of Cambridge, 19 J.J. Thomson Ave., Cambridge, UK 
        \and
        Kavli Institute for Cosmology, University of Cambridge, Madingley Road, Cambridge, UK
    \and
    Instituto de Astrofísica, Facultad de Física, Pontificia Universidad Cat\'olica de Chile, Casilla 306, Santiago 22, Chile
        \and
    Space Telescope Science Institute, 3700 San Martin Drive, Baltimore, MD 21218, USA
    \and
    INAF – Osservatorio Astrofisico di Torino, Via Osservatorio 20, 10025 Pino Torinese, Italy
 }

   \date{Received ; Accepted}

% \abstract{}{}{}{}{} 
% 5 {} token are mandatory
 
  \abstract{
  Quasar feedback in the form of powerful outflows is invoked as a key mechanism to quench star formation in galaxies, although direct observational evidence is still scarce and debated. Here we present Early Release Science JWST NIRSpec IFU observations of the $z$=1.59 prototypical obscured  Active Galactic Nucleus (AGN) XID2028: This target represents a unique test case for studying quasar feedback at the peak epoch of AGN-galaxy co-evolution because extensive multi-wavelength coverage is available and a massive and extended outflow is detected in the ionised and molecular components. 
  With the unprecedented sensitivity and spatial resolution of the JWST, the NIRSpec dataset reveals a wealth of structures in the ionised gas kinematics and morphology that were previously hidden in the seeing-limited ground-based data. In particular, we find evidence of an interaction between the interstellar medium of the galaxy and the quasar-driven outflow and radio jet that produces an expanding bubble from which the fast and extended wind detected in previous observations emerges. The new observations confirm the complex interplay between the AGN jet, wind and the interstellar medium of the host galaxy, highlighting the role of low-luminosity radio jets in AGN feedback. They also clearly show the new window that NIRSpec opens for detailed studies of feedback at high redshift.
  }
 

   \maketitle
%
%-------------------------------------------------------------------

\section{Introduction}

Massive and fast outflows are almost ubiquitous in luminous active galaxies. They are now observed in different gas phases and on different physical scales, from parsec-scale ultrafast ($\sim$10\% of the speed of light) outflows detected in X-rays (e.g. Pounds et al. \citeyear{pounds03}, Nardini et al. \citeyear{nardini15}, Matzeu et al. \citeyear{matzeu23}) to kiloparsec-scale outflows observed in atomic, molecular, and ionised gas with velocities up to $\sim 1000\ \rm km\ s^{-1}$ (e.g. Rupke \& Veilleux \citeyear{rupke11}, Cicone et al. \citeyear{sicone14}, Carniani et al.\citeyear{carniani15}, Harrison et al. \citeyear{harrison16}, F\"orster-Schreiber et al. \citeyear{natascha19}, Kakkad et al. \citeyear{kakkad20}).

Several models have suggested that these massive and fast outflows are able to suppress star formation (SF) activity in the host galaxy by removing and heating the interstellar medium (ISM; e.g. King \citeyear{king05}, Costa et al. \citeyear{costa15}). However, from the observational point of view, it has not yet been clearly assessed whether and how outflows affect the global level of SF in the host galaxies (see e.g. Balmaverde et al, \citeyear{balmaverde16}, Woo et al. \citeyear{woo17}). 

Active galactic nucleus (AGN) feedback is expected to reach its maximum efficiency at $z\sim2$, where both SF and black hole accretion histories peak (e.g. Madau \& Dickinson \citeyear{madau14}). Essentially all massive early-type galaxies are already formed at $z\sim1$ (Cimatti, Daddi \& Renzini \citeyear{cimatti06}), with indications that quenching must have taken place several billion years before (Renzini \citeyear{renzini06}). The impact of AGN-driven outflows critically depends on the ISM properties and gas content, which are different at $z\sim2$ from what is observed in local analogues (e.g. Kewley et al. \citeyear{kewley13}; Steidel et al. \citeyear{steidel14}; Coil et al. \citeyear{coil15}, Tacconi et al. \citeyear{tacconi20}): It is thus fundamental to study AGN outflows at these redshifts.

\begin{figure*}
        \begin{center}
                \includegraphics[width=0.49\textwidth]{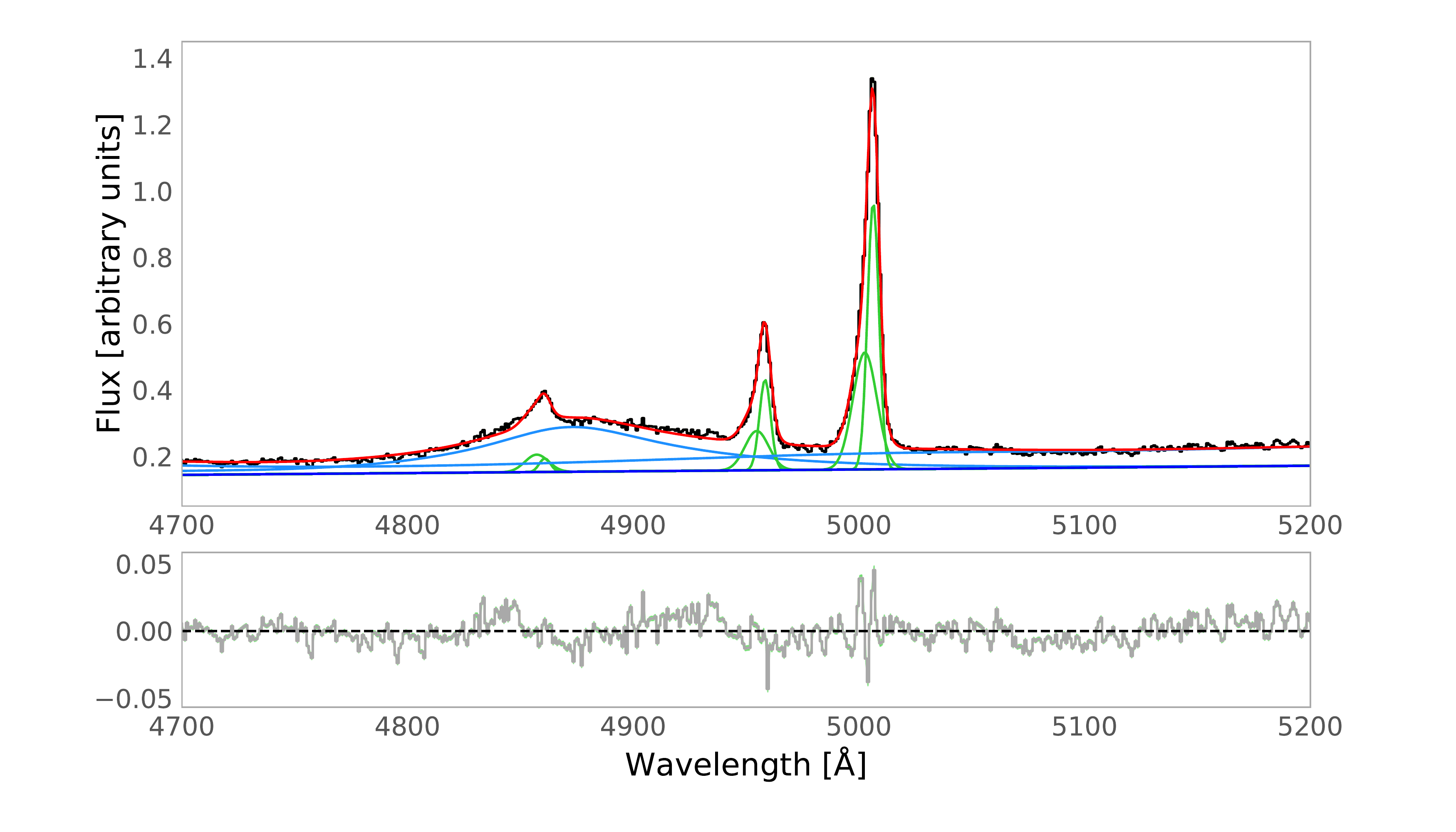}
        \includegraphics[width=0.49\textwidth]{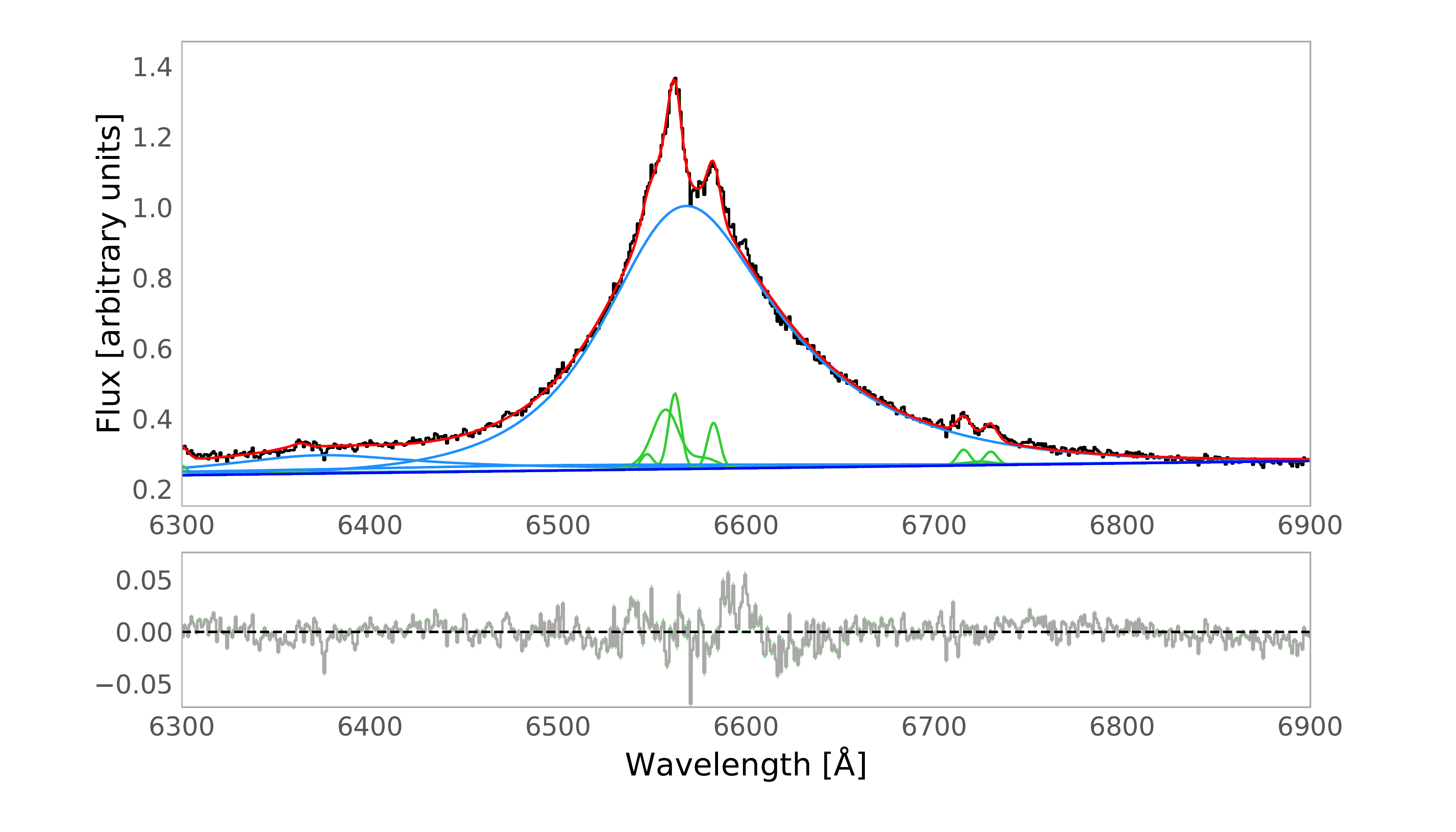}
        \end{center}
        \caption{
        \hb\ (left panel) and \ha\ (right panel) spectral region fit in $5\times5$ NIRSpec spaxels centred on the QSO position. In the upper panels, the data are shown in black, and the best-fitting broad line region components (e.g. \hb, \ha, and FeII) are shown in blue, the narrow line region/host galaxy components (i.e. \hb, [OIII], \ha, [NII], and [SII]) are shown in green, and the total flux is presented in red. The lower panels show the fit residuals.
    } 
        \label{FigBLR}
\end{figure*}

XID2028 ($z$=1.5930), originally discovered in the XMM-COSMOS survey (RA 10:02:11.27 DEC +01:37:06.6; Brusa et al. \citeyear{brusa10}), is one of the best-studied objects in this crucial feedback phase. From an optical point of view, XID2028 is classified as a type 1.8-1.9 quasar on the basis of the significant extinction in the broad line region (BLR)  component as measured from the H$\alpha$/H$\beta$ ratio (see Brusa et al. \citeyear{brusa15a}), and the lack of broad emission lines in the rest-frame UV.  Based on the large multi-wavelength coverage, it has reasonably well-constrained bolometric luminosity $\rm log(L_{AGN})=46.3$, stellar mass $\textrm{M}_*\sim4.5\times10^{11}\ \textrm{M}_{\sun}$ , and SFR $\sim250\ \textrm{M}_{\sun}/\textrm{yr}$, which were derived from fits of the UV to far-IR (FIR) spectral energy distribution (Lusso et al. \citeyear{lusso12}, Brusa et al. \citeyear{brusa15a}, Perna et al. \citeyear{perna15}). 

The presence of a massive ($\rm \dot M \sim 300\ \textrm{M}_{\odot}/\textrm{yr}$) and extended ($\sim 13$ kpc) ionised outflow, traced by \oiiib\ emission, has independently been advocated by X-shooter slit spectroscopy (Brusa et al. \citeyear{brusa15a}, Perna et al. \citeyear{perna15}) and SINFONI\footnote{Spectrograph for INtegral Field Observations in the Near Infrared} IFU observations (Cresci et al. \citeyear{cresci15a}; C15).  The outflow appeared to be located at the centre of a cavity traced by rest-frame U-band HST/ACS imaging and possibly by a narrow \ha\ component attributed to star formation in the host by C15 (but this scenario has been contested  by Scholtz et al. \citeyear{scholtz21}). This suggested that the wind removes the gas from the host galaxy (so-called negative feedback; see e.g. also  Cano-D{\'{\i}}az et al. \citeyear{cano12}, Carniani et al. \citeyear{carniani16},  Cresci \& Maiolino \citeyear{cresci18}), and simultaneously triggers SF by induced pressure at the outflow edges or directly in the wind (so-called positive feedback; see e.g. Silk \citeyear{silk13}, Cresci et al. \citeyear{cresci15b}, Maiolino et al. \citeyear{maiolino17}, Perna et al. \citeyear{perna20}).

Perna et al. (\citeyear{perna15}) also reported the detection of sodium NaID$\lambda\lambda$5890,5896, magnesium MgII$\lambda\lambda$2796,2803, and MgI$\lambda$2853 absorption lines in the X-Shooter optical and near-IR (NIR) spectra of the source. The absorption lines can be ascribed to an outflowing neutral/low-ionisation gas component along the line of sight, with roughly the same velocity shift as the [OIII] emission line.

XID2028 was then observed by Plateau de Bure Interferometer in CO(3-2) (Brusa et al. \citeyear{brusa15b}), and by ALMA in CO(2-1), CO(5-4) and in the 1.3 mm continuum by Brusa et al. (\citeyear{brusa18}; B18). They derived a total molecular gas mass of $\sim10^{10}\ \textrm{M}_{\odot}$, which corresponds to a very low gas fraction ($<$5\%) and depletion times (i.e. the timescale during which all the gas reservoir is converted into stars) of only $40-75$ Myr, supporting a scenario in which outflows affect the gas reservoirs. While the bulk of the CO emission appeared to be located in a rotating disk in the inner few kiloparsec, B18 also detected extended high-velocity gas, which they interpreted as a signature of a galaxy-scale molecular outflow that is spatially coincident with the blueshifted ionised gas outflow. A possible redshifted counterpart of the outflow was also detected on the opposite side of the galaxy (see Fig.~\ref{CO_VLA}). Summing up both the ionised and molecular gas components, B18 estimated a total mass rate of the outflowing gas of $\rm \dot M_{\textrm{out,tot}} \sim 500-800\ \textrm{M}_{\odot}/\textrm{yr}$.

%Finally, Vardoulaki et al. (\citeyear{vardoulaki19}) presented radio imaging of XID2028 among the multi-component radio sources identified in the VLA-COSMOS Large Project at 3 GHz (0.75" resolution, 2.3 $\mu \textrm{Jy}\ \textrm{beam}^{-1}$ RMS). The galaxy is identified as \#10965 in their catalogue but surprisingly classified as a Star-Forming Galaxy based on the radio morphology, despite the two blobs extending in opposite directions from the centre, at locations corresponding to the blue and red-shifted outflows detected in the ionised and molecular data.

The available numerous multi-wavelength observations and its unique properties therefore make XID2028  one of the best objects for characterising outflows and feedback at z$\sim$1-2. It was thus selected to be observed with the Near Infrared Spectrograph (NIRSpec) and Mid-Infrared Instrument (MIRI) integral field units (IFU) on board the James Webb Space Telescope (JWST) during Early Release Science (ERS) observations. In this paper, we present the NIRSpec dataset, which reveals a wealth of structures that were hidden in the previous seeing-limited ground-based data.

The paper is organised as follows: in Sect.~\ref{data} we present the observations, data reduction, and analysis. The results are presented in Sect.~\ref{results}, and we draw our conclusions in Sect.~\ref{conclusions}. Throughout the paper, we adopt the cosmological parameters $\textrm{H}_0 = 70\ \textrm{km}\ \textrm{s}^{-1}\ \textrm{Mpc}^{-1}$, $\Omega_{\textrm{m}}=0.3,$ and $\Omega_{\lambda}=0.7$ (Spergel et al. \citeyear{spergel03}).

\section{NIRSpec observations, data reduction, and analysis} \label{data}

\subsection{Observations}

XID2028 was observed with NIRSpec (Jakobsen et al. \citeyear{jakobsen22}, B{\"o}ker et al. \citeyear{boker22}) on board the JWST on 20 November 2022 as part of the  Q-3D ERS Proposal 1335 (PI D. Wylezalek). The program aims at studying the properties of three well-studied and luminous quasars (QSOs) with the unprecedented sensitivity, spatial resolution, and spectral coverage provided by the JWST (Wylezalek et al. \citeyear{wylazalek22}).
% NOT NEEDED, MIRI is not even mentioned above. This work focuses on the NIRSpec IFU observations of XID2028, while the analysis of the MIRI observations will be presented in a future paper. 

The NIRSpec IFU observations provide spatially resolved imaging spectroscopy in the wavelength range $0.97-1.82\ \mu m$ at $\textrm{R}\sim2700$ with the G140H/F100LP grating/filter pair over a $3"\times 3"$ field of view with $0.1" \times 0.1"$ spatial elements. The FWHM of the NIRSpec PSF is $\sim45$ mas at 1.3 $\mu m$, which is the observed wavelength of \oiiib\ emission for this source.
The observations were taken with an NRSIRS2 readout pattern with 16 groups, using a nine-point medium-cycling dither pattern. This resulted in a total exposure time of 2.95 hours.
\begin{figure}
        \begin{center}
                \includegraphics[width=0.49\textwidth]{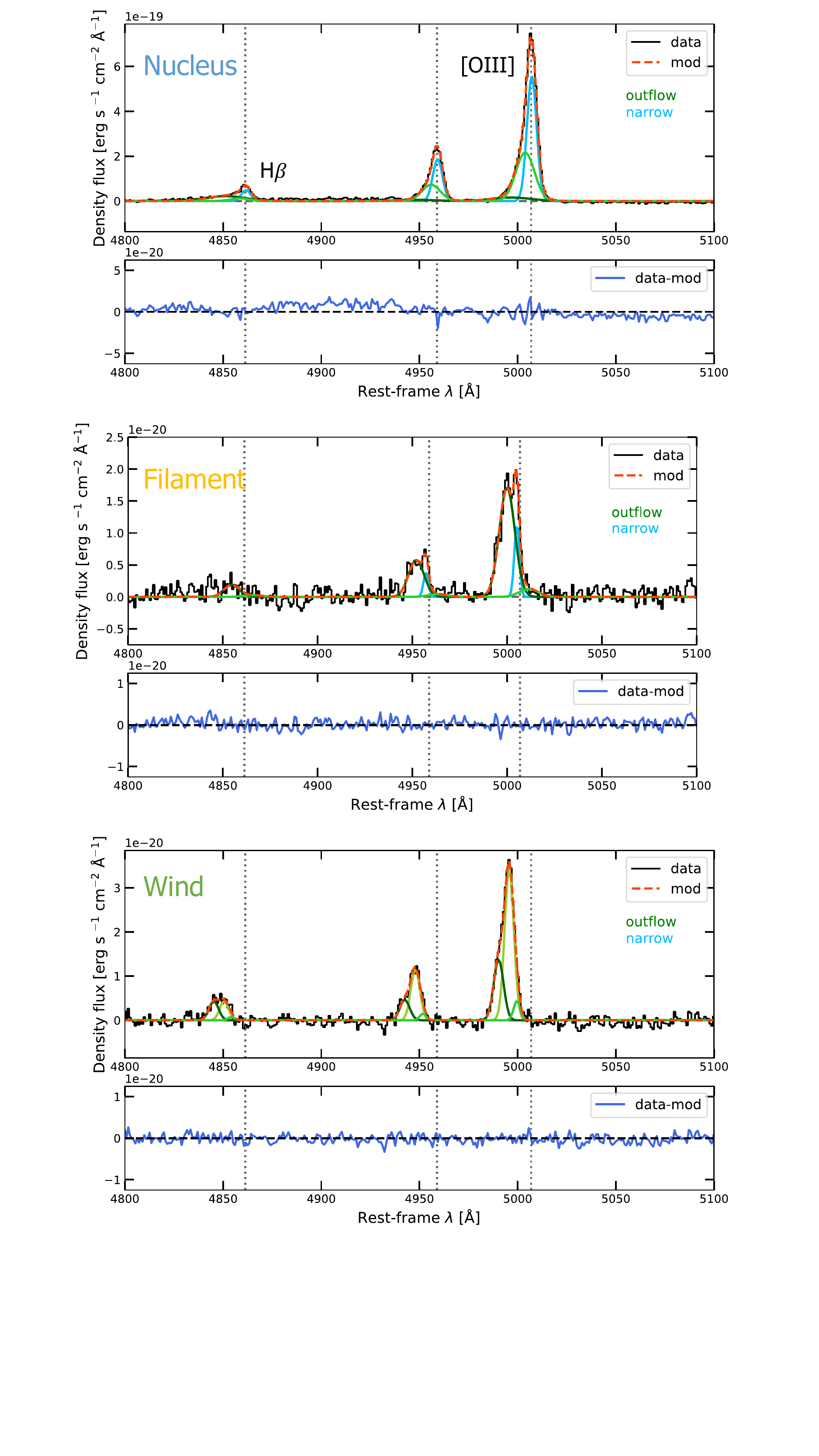}
        \end{center}
        \caption{Representative examples of our multi-component fitting of \hb\ and [OIII] emission lines in single spaxels from three different regions of the galaxy, as marked with coloured crosses in Fig.~\ref{channels}. From the top, the panels refer to a nuclear region spaxel, to a filament spaxel (see text), and to a spaxel in the most blueshifted wind region, respectively. The NIRSpec data are shown in black, the best-fitting systemic narrow components are shown in cyan, and the Gaussian components used to reproduce the outflow are shown with different tones of green. 
 We classified the Gaussian components in each spaxel with a velocity shift |v|<300 km/s (with respect to the systemic velocity of the QSO, marked with a dotted line) and $\sigma< 300$ km/s as narrow (see Sect. \ref{disk} and \ref{ionout}). Systemic narrow line emission dominates in the nucleus and becomes fainter along the filaments, while only outflow emission is detected in the wind region (see Sect.~\ref{gaskin}). Residuals of the emission-line fitting are shown in blue in a sub-panel for all regions.
    } 
        \label{samplespec}
\end{figure}
\begin{figure*}
        \begin{center}
                \includegraphics[width=0.9\textwidth]{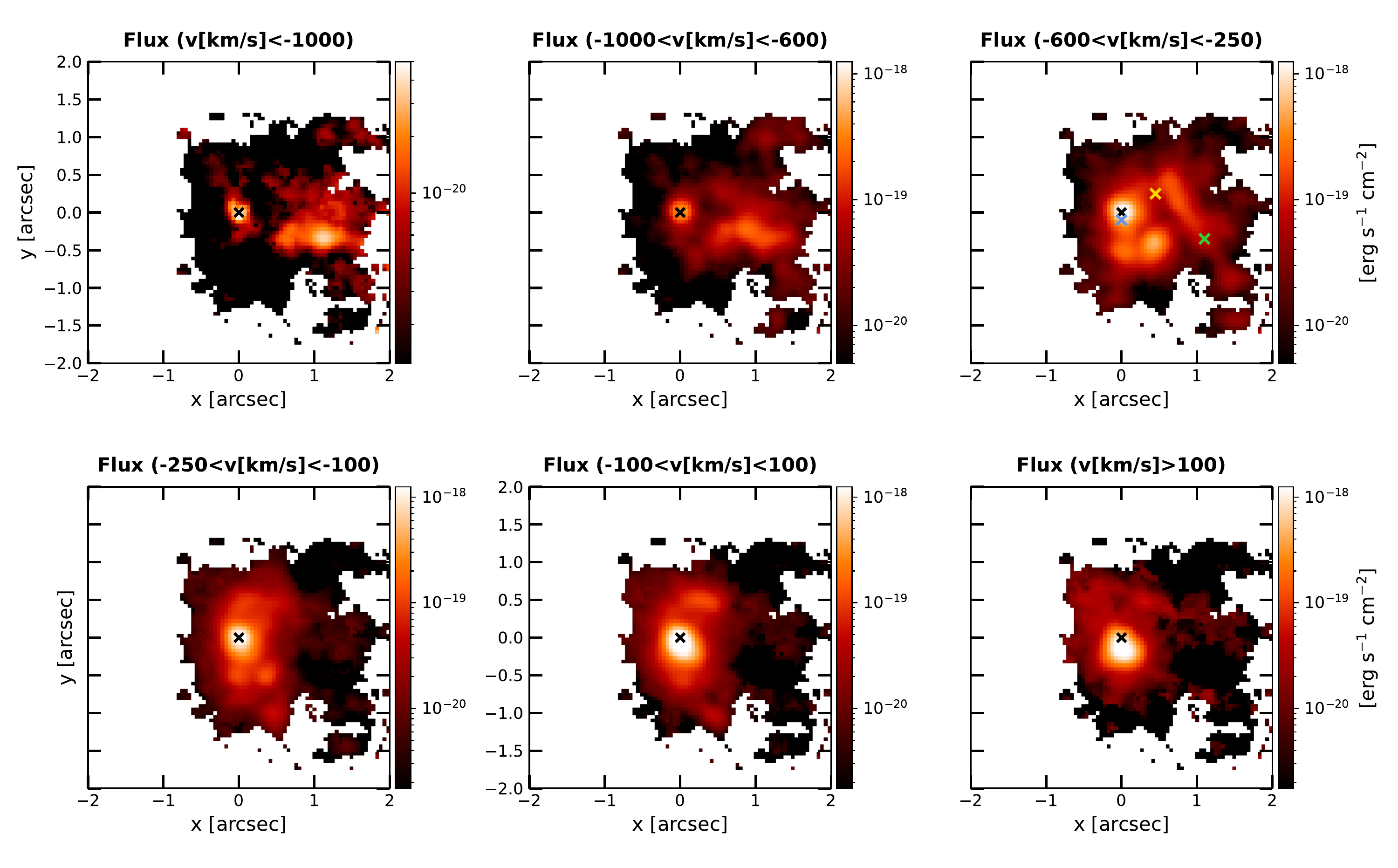}
        \end{center}
        \caption{
        Channel maps for the \oiiib\ line emission. The six panels show the [OIII] line flux in different velocity bins. In the upper left panel, the highest blueshifted velocities are shown, highlighting the fastest part of the wind. In the following panels, the filaments connecting the outflow to the quasar location become evident. In contrast, the lower right panel shows the possible redshifted outflow emission to the north-east. The maps show the spaxel with S/N$>$3 on the total [OIII] flux. At the distance of the target, the scale is $\sim8.5$ kpc/". The coloured crosses show the location of the extraction spaxels for the spectra shown in Fig.~\ref{samplespec}.
    } 
        \label{channels}
\end{figure*}

\subsection{Data reduction}

The data reduction was performed with the JWST calibration pipeline version 1.8.2, using the context file {\it jwst\_1014.pmap}. The individual raw images were first processed for detector-level corrections using the {\it Detector1Pipeline} module of the pipeline (stage1). Then, the resulting count-rate images were calibrated using the {\it Calwebb\_spec2} module (stage2). At this stage, WCS-correction, flat-fielding, and flux calibrations were applied to convert the data from units of count rate into flux density. At the time of writing, many NIRSpec calibration files contain placeholder values: for this reason, no proper flat-fielding has been applied to the data. This has implications for the shape of the background continuum emission, but does not affect the analysis of the bright line emission from the QSO.
The individual stage2 images were then resampled and coadded onto a final data cube through the {\it Calwebb\_spec3} processing (stage3). We also applied a number of additional steps (and custom modifications of the current pipeline code) to improve the data reduction quality. Moreover,  different configurations were also used to obtain additional data products and test the pipeline robustness (e.g. of flux and spatial resolution recovery). In particular:
\begin{itemize}
    \item We applied the snowball flagging for the {\it jump} step during the first stage of the pipeline. This removes data artefacts caused by large cosmic-ray impacts on the detector.
    \item We further processed each individual count-rate frame to correct for varying zero-level in the dithered frames at the end of stage1. The median value (computed considering the entire image) was subtracted from each frame to obtain a base level consistent with zero counts per second.
    \item The obtained count-rate images were also processed to remove the $1/f$ noise pattern, a low-frequency noise for which the noise power is inversely proportional to the frequency. This appears as correlated vertical noise, which we modelled in each column (i.e. along the spatial axis)  with a low-order polynomial function after flagging and removing all bright pixels (e.g. associated with the observed target) through a sigma-clipping algorithm. The modelled $1/f$ noise was then subtracted before proceeding with stage2 of the pipeline.
    \item The flux calibration was performed using %two different approaches: the first one is obtained using the {\it photom} step of Stage2; instead, the second approach takes advantage of 
    the commissioning observations of the standard star TYC 4433-1800-1 (PID 1128, o009). %In the latter case, the flux calibration is performed as a post-processing correction: 
    We reduced the star observations with the same pipeline version and context file as the science frames, and we used it to derive the response curve of the instrument required to convert count rates into flux densities.
    \item The {\it outlier\_detection} step in stage3 was used to identify and remove the numerous remaining spikes from the datacubes that are due to cosmic rays and other artefacts and that were not corrected for in the previous reduction steps. Unfortunately, this step cannot be used in the current version of the pipeline, as it identifies too many false positives and seriously compromises the data quality by removing too many spaxels with reliable signal. We therefore used a post-processing correction based on a $\sigma$-clipping cut to exclude all spikes in the reduced datacubes (at the spaxel level). 
    \item We applied the {\it cube\_build} step to produce a final datacube with a spaxel size of $0.05"$, obtained with the drizzle weighting. In this step, we patched the {\it cube\_build} script by fixing a bug affecting the drizzle algorithm, as implemented in version 1.8.5\footnote{{\it cube\_build} code changes in \url{https://github.com/spacetelescope/jwst/pull/7306}}.
    \item As the final astrometric solution calculated by the pipeline resulted in an offset of $\sim0.1"$ of the QSO position, we corrected the WCS using the archival HST/ACS imaging in the F814W filter as reference (Koekemoer et al. \citeyear{koekemoer07}, C15) to align the bright QSO nucleus in NIRSpec with its expected HST position. 
    %\item We applied the {\it cube\_build} step to produce two combined data cubes: one with a spaxel size of $0.1"$, obtained with the \textit{emsm} weighting (with higher signal-to-noise at spaxel level, and less affected by wiggles in the continuum), and a second one with a spaxel size of $0.05"$, obtained with the drizzle weighting (with higher spatial resolution).
\end{itemize}

\subsection{Data analysis}

For the spectral analysis of the data, we adopted the fitting code presented in Marasco et al. (\citeyear{marasco20}, \citeyear{marasco22}) and Tozzi et al. (\citeyear{tozzi21}) to analyse IFU data of local and high-z AGN and star-forming galaxies. 
The first step of the data analysis models the unresolved BLR. We fit a BLR model to a nuclear spectrum extracted in a $5\times5$ spaxel ($0.25"\times0.25"$) region around the AGN location. Following C15, we fitted the asymmetric broad lines (\ha, \hb, \hy\ and \heii) with a broken power-law distribution convolved with a Gaussian. The broad line centroids,  power-law indices, and widths were tied to be the same for all lines, with the exception of the \ha\ Gaussian width, which required a slightly different width to reproduce the observed spectrum. The FeII  emission was reproduced using a combination of different CLOUDY (Ferland et al. \citeyear{ferland98}) models. Two additional BLR Gaussian components were added at $\sim-8000$ and $\sim-16000$ km $\textrm{s}^{-1}$ from \ha\ (see e.g. Carniani et al. \citeyear{carniani16}). The forbidden lines and the narrower components of the BLR lines were fitted with two Gaussian components each, forcing each of them to have the same velocity and width for all species. The resulting fit is shown in Fig.~\ref{FigBLR} for the spectral regions around \hb\ and \ha.

We used the derived BLR model to subtract its unresolved emission from the data spaxel by spaxel. We used the pPXF code (Cappellari \citeyear{cappellari17}) to fit the entire datacube, allowing the BLR template to vary only in flux, as expected for unresolved emission that is distributed as the instrumental point spread function (PSF). We additionally modelled the residual continuum with a first-order polynomial and the extended line emission of the main species as combinations of multiple Gaussian components, ranging from one to three. A Kolmogorov-Smirnov test on the residuals was used to select the optimal number of components in each spaxel to reproduce the observed profile while minimising the number of free parameters (see Marasco et al. \citeyear{marasco20}). The rescaled best-fitting BLR template and continuum models were then subtracted spaxel by spaxel, obtaining a BLR-subtracted cube containing only the extended line emission due to the NLR, the host galaxy, and the outflow components. 

We then refit the BLR-subtracted cube to carefully model the residual line emission. We fixed the ratio of the lines in each of the \oiii\ and \nii\ doublets to the theoretical value of 3, and we forced each Gaussian component of all emission lines to have the same line profile shape (centroid velocity and velocity dispersion), assuming that they are emitted from regions with similar kinematics. This reduced the degeneracy in the fitting procedure, especially for the \ha\ and [NII] multiplet in spaxels with complex kinematics. As before, we performed the fit with up to three Gaussian components for each line, and we selected the optimal number of components to reproduce the line profile spaxel by spaxel. Three examples of multi-component line fitting in different regions of the galaxy are shown in Fig.~\ref{samplespec}. Their locations are marked with coloured crosses in Fig.~\ref{channels}.

\begin{figure*}
        \begin{center}
                \includegraphics[width=0.9\textwidth]{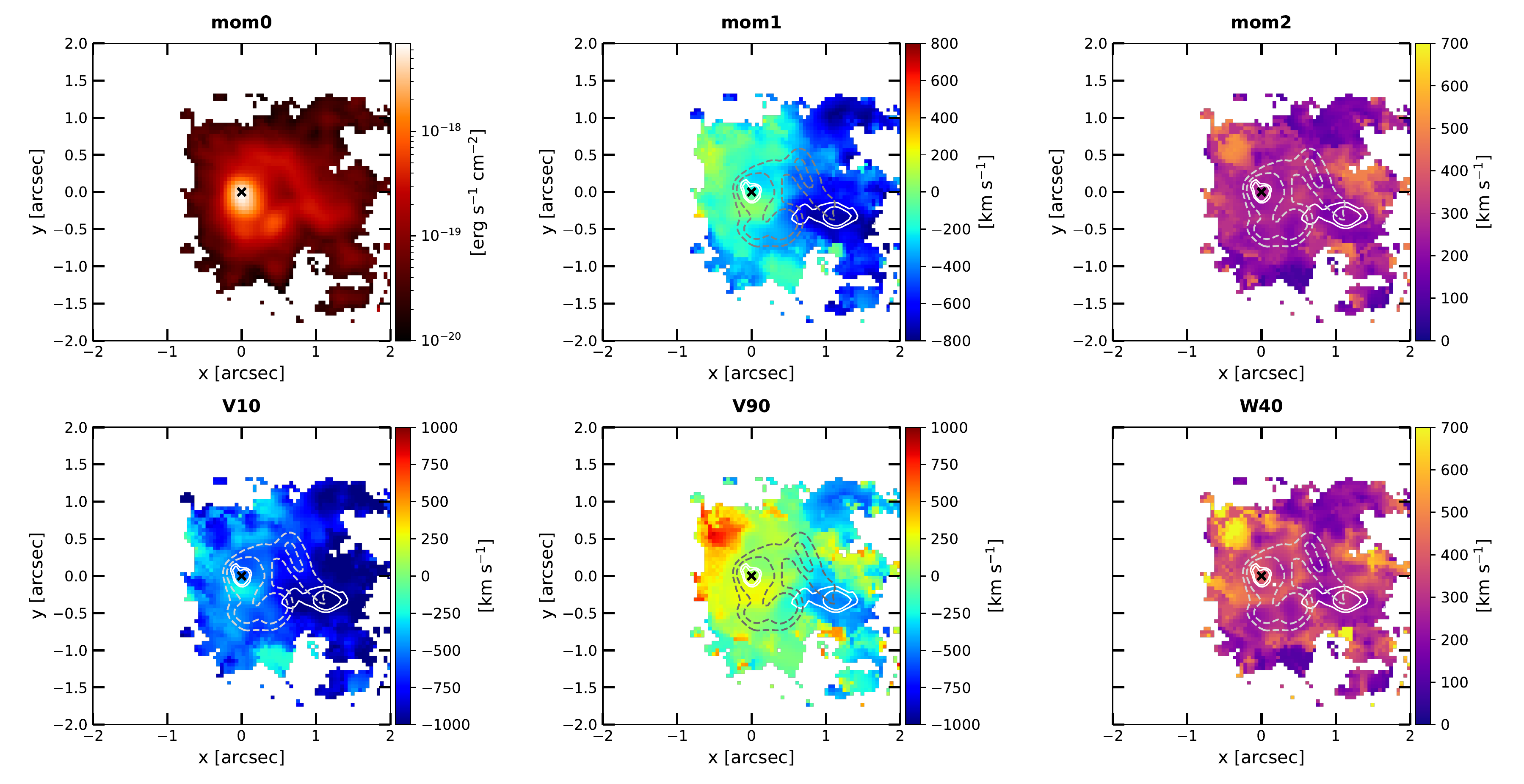}
        \end{center}
        \caption{
        Kinematics of the [OIII] line emission. The upper panels show moment 0, moment 1, and moment 2 of the whole line profile velocity. The lower panels instead show $\textrm{v}_{10}$, the velocity at the 10th percentile of the overall emission-line profile in each spaxel, $\textrm{v}_{90}$, the 90th percentile, and $\textrm{W}_{40}$, i.e., $\textrm{W}_{40} = \textrm{v}_{50} - \textrm{v}_{10}$, the line width containing 40\% of the emission line flux. Solid contours represent 30\% and 50\% of the peak emission in the bluest channel map ($\textrm{v}<-1000$ km/s) in Fig.~\ref{channels}, and the dashed contours the 5\% and 10\% emission of the third channel ($-600<\textrm{v}<-250$ km/s). The maps show the spaxel with S/N$>$3 on the total [OIII] flux, and the cross marks the position of the QSO.
    } 
        \label{kinmaps}
\end{figure*}

\section{Results} \label{results}

\subsection{Bubbles and outflows: Extended ionised gas kinematics} \label{gaskin}

%In the following, we will mostly focus on the kinematics of the \oiiib\ line, which is an ideal tracer of extended outflowing ionized gas, as it cannot be produced in the high-density, sub-parsec scales typical of the BLR. 

Fig.~\ref{channels} shows channel maps of the modelled [OIII] line flux in different velocity bins. The same pattern is also present in H$\alpha$, H$\beta,$ and the other emission lines, but we present a kinematic analysis of [OIII] here because this bright forbidden line is not contaminated by emission from the BLR and is thus an ideal tracer of outflows. Prominent collimated blue emission is evident in the most blueshifted velocity bin ($\textrm{v}<-1000$ km/s). This highly blueshifted emission is located in the same region as the [OIII] outflow detected in earlier SINFONI data by C15, as well as the blueshifted molecular outflow discovered by B18. Notably, as seen also in the SINFONI data, the blue [OIII] outflow emission peaks at a projected distance of $\sim 0.7"$ ($\sim6$ kpc) west of the QSO. 

At lower blueshifted velocity ($-600<\textrm{v}<-250$ km/s), a cavity of suppressed line emission becomes evident between the QSO and the outflow, surrounded by bright [OIII] filaments that seem to connect the QSO with the region with the fastest outflow emission. This filamentary structure around the cavity roughly corresponds to the elongated regions detected in HST/ACS imaging and narrow \ha\ in C15. 

%Finally, fast, receding line emission is detected towards the northeast in the reddest channels ($\textrm{v}>100$ km/s). ({\bf NOT SHOWN IN FIG. 2 THESE CHANNELS}.) This emission spatially corresponds to the red-shifted molecular outflow discussed in B18, and thus might represent the receding ionised outflow undetected in the earlier SINFONI data. {\bf MB: you cannot refer here to the upper panel of fig. 3 to illsutrate this, because the upper panel of Fig. 3 shows only the blue channels at v$<-300$ km/s while here you are commenting red channels and the superposition with the red molecular outflow. SUGGESTION: put the blue molecular outflow contours in the first and/or second panel in fig. 2 and the red molecular outflow conturs in the last panel of Fig. 2. Remove therefore the upper panel of Fig. 3 from the figure and move this figure in the relevant session before Fig. 8}

The kinematics maps in Fig.~\ref{kinmaps} also highlight the features discussed above. %Given the complexity of the line profile, we first adopted a non-parametric definition to characterise the widths and velocities within each spaxel of the NIRSpec datacube. 
The figure shows the moment~0, moment~1, and moment~2 maps of the total \oiiib\ emission line velocity in the upper panels, with superimposed contours of the moment 0 maps. The bottom panels show $\textrm{v}_{10}$, the velocity at the 10th percentile of the overall emission-line profile in each spaxel, $\textrm{v}_{90}$, the 90th percentile, and $\textrm{W}_{40}$, defined as the line width containing 40\% of the emission line flux $\textrm{W}_{40} = \textrm{v}_{50} - \textrm{v}_{10}$. In this parametrisation, v$_{10}$ traces the highest-velocity blueshifted gas in the data, with velocities as high as $\textrm{v}_{10} \sim -1000$ km/s towards our line of sight. These velocities are not compatible with either rotational motions in the host, given their high values and their direction perpendicular to the velocity gradient in the host (see B18), or with star formation-driven outflows, which are generally characterised by lower peak velocities (see e.g. Arribas et al. \citeyear{arribas14}, Fluetsch et al. \citeyear{fluetsch19}, F{\"o}rster Schreiber et al. \citeyear{natascha19}). We also exclude the possibility that they may be caused by a companion galaxy because the line width at the highest blueshifted velocities is $\sigma_{\textrm{blue}}\sim450$ km/s, which would imply a very high-mass companion galaxy that is undetected in continuum\footnote{A continuum feature, located nearby ($\sim0.3"$) but not coincident with the blueshifted emission, is in fact detected only in one of the two detectors covering the spectra at that location, at wavelengths between 1.465 $\mu m$ and 1.89 $\mu m$, while it suddenly disappears at wavelengths $\lambda< 1.43\ \mu m$ in the other detector. Because the data were not flat-field corrected for the differential pixel-to-pixel variations because a reliable flat-field calibration is not yet available for NIRSpec, and because multiple similar continuum structures were present in the datacube with different spatial positions in wavelength regions covered by different detectors, we attribute this feature to pixel-to-pixel continuum level variations and consider the interpretation as a companion galaxy very unlikely.}, nor is it dected in deep ACS imaging of the source (see C15). 
\begin{figure}
        \begin{center}
                \includegraphics[width=0.49\textwidth]{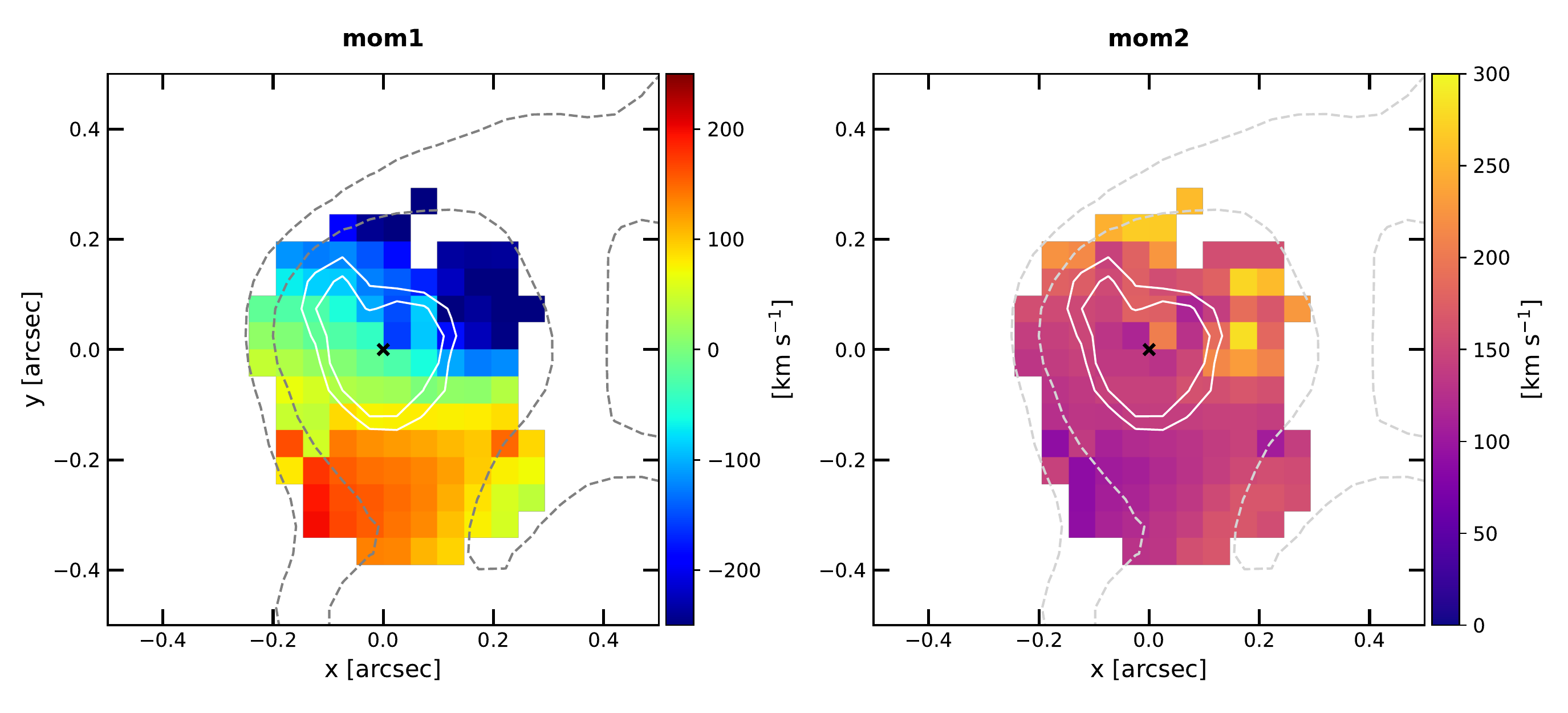}
        \end{center}
        \caption{Moment 1 and moment 2 maps for the component of the \oiiib\ line emission tracing gas in the host galaxy (see text). The contours are the same as in Fig.~\ref{kinmaps}. An S/N$>$40 threshold has been used to mask lower S/N spaxels and isolate the rotation in the central spaxels. A rotational pattern is suggested by the data that is compatible with the CO(5-4) rotation detected by B18. 
    } 
        \label{rotation}
\end{figure}

Similarly, $\textrm{v}_{90}$ traces the highest redshifted velocities, up to v$_{90}\sim 800$ km/s. These are located towards the north-east, opposite to the QSO location with respect to the blueshifted velocities, suggesting that the gas is outflowing in opposite directions from the central source, as predicted in bi-conical outflow models. This emission spatially corresponds to the redshifted molecular outflow discussed in B18 (see Fig.~\ref{kinmaps} and Fig.~\ref{CO_VLA}, upper panel), and thus might represent the receding ionised outflow that was undetected in the earlier SINFONI data. The location of the redshifted outflows also shows the highest line widths, with W$_{40}\sim 700$ km/s. 

The moment 0 [OIII] map also shows the filamentary structure that was identified in the channel maps, which spatially and in velocity connects the QSO to the collimated blueshifted outflow. As shown in the moment 1 map, the filaments show a symmetric intermediate velocity of $\sim-300$ km/s between the QSO and the outflow (see Fig.~\ref{samplespec}), and their velocity gradient shows increasing blueshifted velocities towards the position of the collimated outflow. This suggests that the two are physically connected in some way.

Highly blueshifted emission, but with fainter [OIII] emission, is also detected towards the northwest, with v$_{10}$ velocities comparable to those observed in the outflow. The detection of a possible continuum source in the NIRSpec datacube at that location may suggest a companion galaxy. Reliable flat-field calibration for NIRSpec is required to test this hypothesis, but is not yet available, however.

\subsection{Possible rotating disk in the host galaxy} \label{disk}

We also attempted to measure the gas kinematics in the disk of the host galaxy. To do this, we selected the  Gaussian component with the highest peak in each spaxel, excluding those with a velocity shift $|\textrm{v}|>300$ km/s or with $\sigma>300$ km/s. We show the resulting moment 1 and moment 2 maps in Fig.~\ref{rotation}, where we impose a cut of S/N$>$40 in the displayed spaxels to better isolate the rotation-like pattern in the central region. 
Interestingly, the velocity gradient obtained in this way is consistent with the one revealed by B18 in CO(5-4), which was interpreted as a compact rotating disk in the host galaxy. In particular, both the position angle PA $\sim125^{\circ}$ and the magnitude of the velocity gradient ($-200 \lesssim \textrm{v} \lesssim 200$ km/s) are consistent with the molecular gas measurements. Assuming that this velocity gradient traces the major axis of the host galaxy, this orientates the red- and blueshifted outflow components roughly along the minor axis. Assuming an inclination of 30$^{\circ}$, B18 derived a dynamical mass $\textrm{M}_{dyn}\sim6-8 \times 10^{10}\ \textrm{M}_{\odot}$ within the region in which CO is detected (r$\sim0.25"$), using both the Cresci et al. (\citeyear{cresci09}) and the Di Teodoro \& Fraternali (\citeyear{barolo15}) kinematical fitting codes.
\begin{figure}
        \begin{center}
                \includegraphics[width=0.5\textwidth]{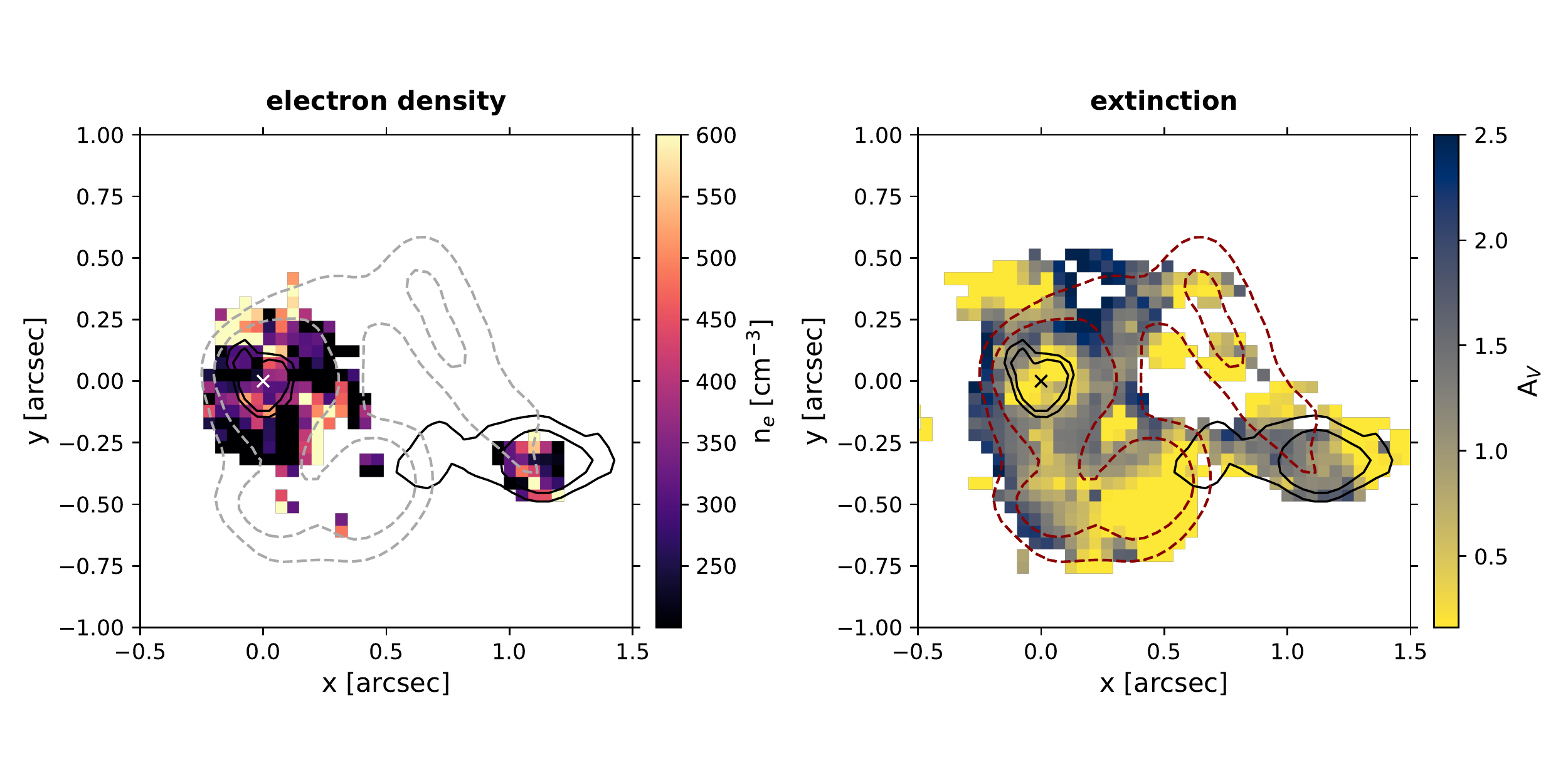}
        \end{center}
        \caption{Electron density map from \siiab\ line ratio (left panel) and extinction map from the Balmer decrement \ha/\hb\ (right panel). The contours used in Fig.~\ref{kinmaps} are overplotted. Both maps display only the spaxels with at least S/N$>$3 in all the emission lines involved, and the QSO location is marked with a cross.
    } 
        \label{avne}
\end{figure}
\begin{figure}
        \begin{center}
                \includegraphics[width=0.5\textwidth]{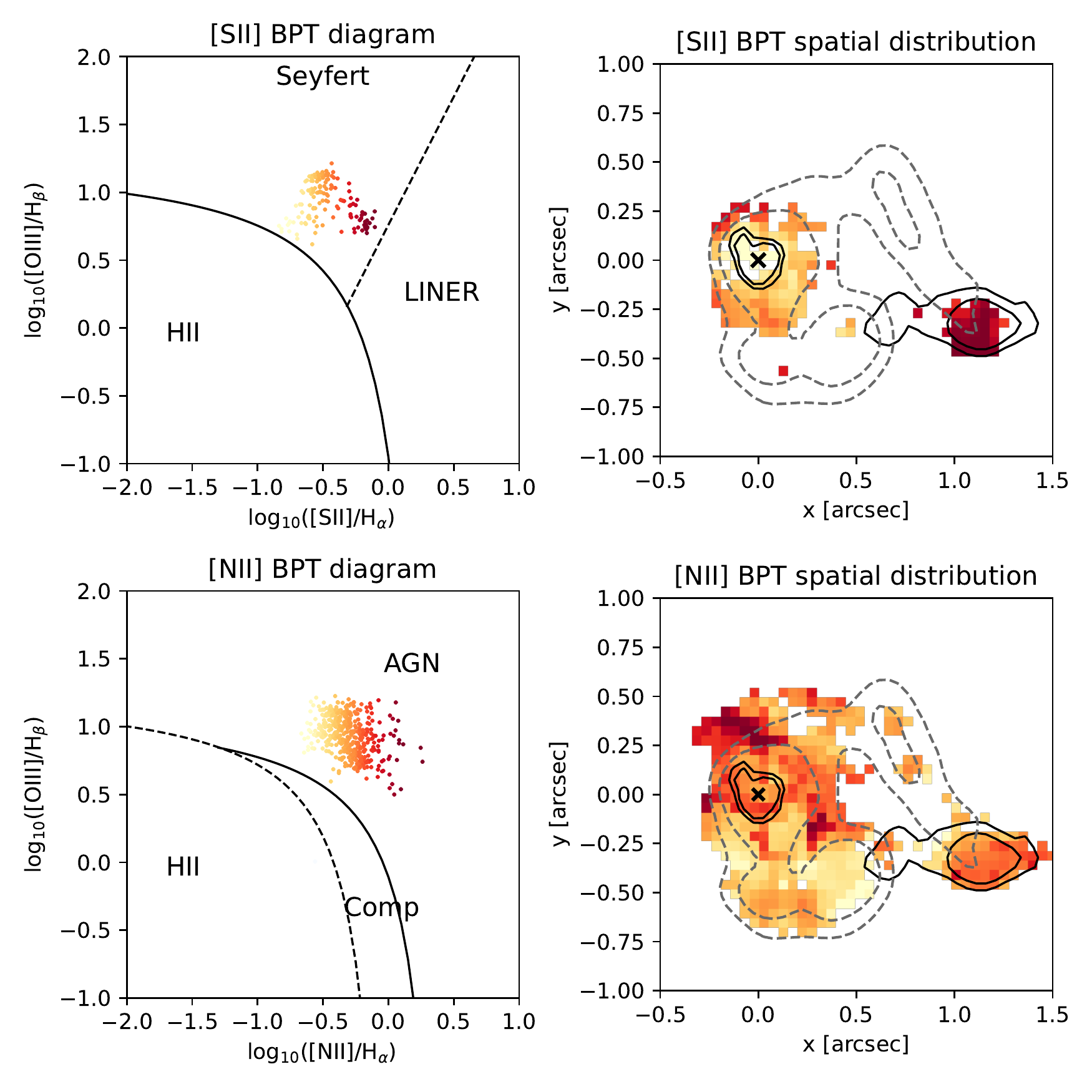}
        \end{center}
        \caption{Resolved [SII]-BPT (upper panels) and [NII]-BPT (lower panels) diagrams for each spaxel with S/N$>$3 in each line. The data points in the BPT diagrams (left panels) are colour coded as a function of their [SII]/\ha\ and [NII]/\ha\ ratios for the [SII]-BPT and [NII]-BPT, respectively. The same colours are used in the corresponding maps (right panels), where the contours shown in Fig.~\ref{kinmaps} are overplotted. The QSO location is marked with a black cross.
    } 
        \label{bpt}
\end{figure}

\subsection{ISM properties} \label{ism}

We explored the physical conditions of the ISM using the observed line ratios in the NIRSpec datacube. 
The electron density n$_e$ was derived using the \siiab\ line ratio. The \siiab\ lines were fitted using the same components used for the brighter forbidden lines, although the deblending into different components is very uncertain due to the low S/N of the broader components. We therefore used the total \siiab\ line flux, the parametrisation of Sanders et al. (\citeyear{sanders16}) and assumed T$_e$=$10^4$ K. In the left panel of Fig.~\ref{avne}, we show the resulting $\rm n_e$ map, which is noisy because the [SII] line emission is faint and the modelling of the two distinct components of the doublet may have degeneracy issues. However, the density seems uniform across the sampled regions where the [SII] lines are detected with S/N$>$3, with values between $200-500$ cm$^{-3}$. This is compatible with similar measurements at $z\sim2$ in star-forming and active galaxies (see e.g. Sanders et al. \citeyear{sanders16}, Masters et al.~\citeyear{masters14}, Steidel et al.~\citeyear{steidel14}, Kaasinen et al. \citeyear{kaasinen17}, Kashino et al. \citeyear{kashino17}). Although a higher density has been found in the outflow component than in the disk component in the hosts of local and high-z AGNs (see e.g. Mingozzi et al. \citeyear{mingozzi19},  F{\"o}rster Schreiber et al. \citeyear{natascha19}), we derive similar values in the central region dominated by the QSO and on the outflow region within the errors. In the outflow region, where the entire emission is blueshifted and samples the outflow, we measured a median of n$_e\sim360\pm180$ cm$^{-3}$, which is consistent with the value derived by an integrated spectrum centred on the outflow region with a radius of 3 spaxels. 

We used the Balmer decrement, \ha/\hb, to derive the dust attenuation map, assuming a Calzetti et al. (\citeyear{calzetti00}) attenuation law and a temperature of $10^4$ K. The map we obtained is shown in the right panel of Fig.~\ref{avne}. The optical extinction seems to decrease towards the nucleus despite the red colours of the QSO. The reason probably is that the rest frame optical lines cannot probe deep in the highly embedded regions around the QSO (see e.g. Cresci et al. \citeyear{cresci10}, Cresci et al. \citeyear{cresci17}). MIRI observations will be useful to probe the dust that is located deeper in this region. Some unsubtracted residuals of the BLR in H$\beta$ might also contribute. The extinction also decreases along the filaments, where shocks might contribute to destroying the dust. A ratio \ha/\hb$\sim4$\ is measured  in the highly blueshifted outflow region, corresponding to a median extinction $\textrm{A}_{\textrm{V}}=1.1\pm0.5$, where the error is given by the standard deviation of $\textrm{A}_{\textrm{V}}$ over the outflow region. 

We also used the so-called BPT diagrams (Baldwin et al. \citeyear{baldwin81}, Kewley et al. \citeyear{kewley06}, Lamareille et al. \citeyear{lamareille10}, Law et al. \citeyear{law21}) to investigate the dominant ionisation source in the different regions of the galaxy. As shown in Fig.~\ref{bpt}, all spaxels are AGN dominated according to [SII]-BPT (\siiab/\ha\ vs \oiiib/\hb) and [NII]-BPT (\niib/\ha\ vs \oiiib/\hb) relative to the full line emission. 
We find that the low-ionisation lines [SII] and [NII] are enhanced along the outflow and in a cavity surrounded by the filaments (see Sect.~\ref{jetism}), possibly by shocks. %, and that the ionisation parameter traced by [OIII]/\hb\ is decreasing at increasing distance from the QSO. 
The BPT diagrams shown in Fig.~\ref{bpt} are relative to the full line profile after removing the BLR component.

By inspecting fit residuals, we searched for a residual narrow \ha\ component tracing star formation in the host galaxy. In previous SINFONI data, C15 detected such an SF component in two distinct spots around the outflow location, where NIRSpec has now unveiled gas filaments.
The component was later not detected in the different analysis performed by Scholtz et al. (\citeyear{scholtz21}). As the residuals of the \ha\ fit are critically dependent on the assumptions made to perform the fit and the BLR subtraction, we fixed the parameters of the fitting of the BLR-subtracted \ha\ to the [OIII] best-fitting parameters to minimise the degeneracy. 
Most of the (small) residuals in the \ha\ flux after subtracting the best fit are detected in the wings of the line, and we do not identify them as additional kinematic components, but rather with an imperfect parametrisation of the BLR wings. Nevertheless, the elongated filaments observed in the rest-frame U band HST/ACS imaging still suggest that the presence of an outflow and an expanding bubble of hot gas, as discussed in the next section, might have a significant effect in triggering star formation and compressing the ISM at their edges. The related \ha\ emission might be difficult to disentangle from the bright QSO contribution, however. 

\begin{figure}
        \begin{center}
                \includegraphics[width=0.45\textwidth]{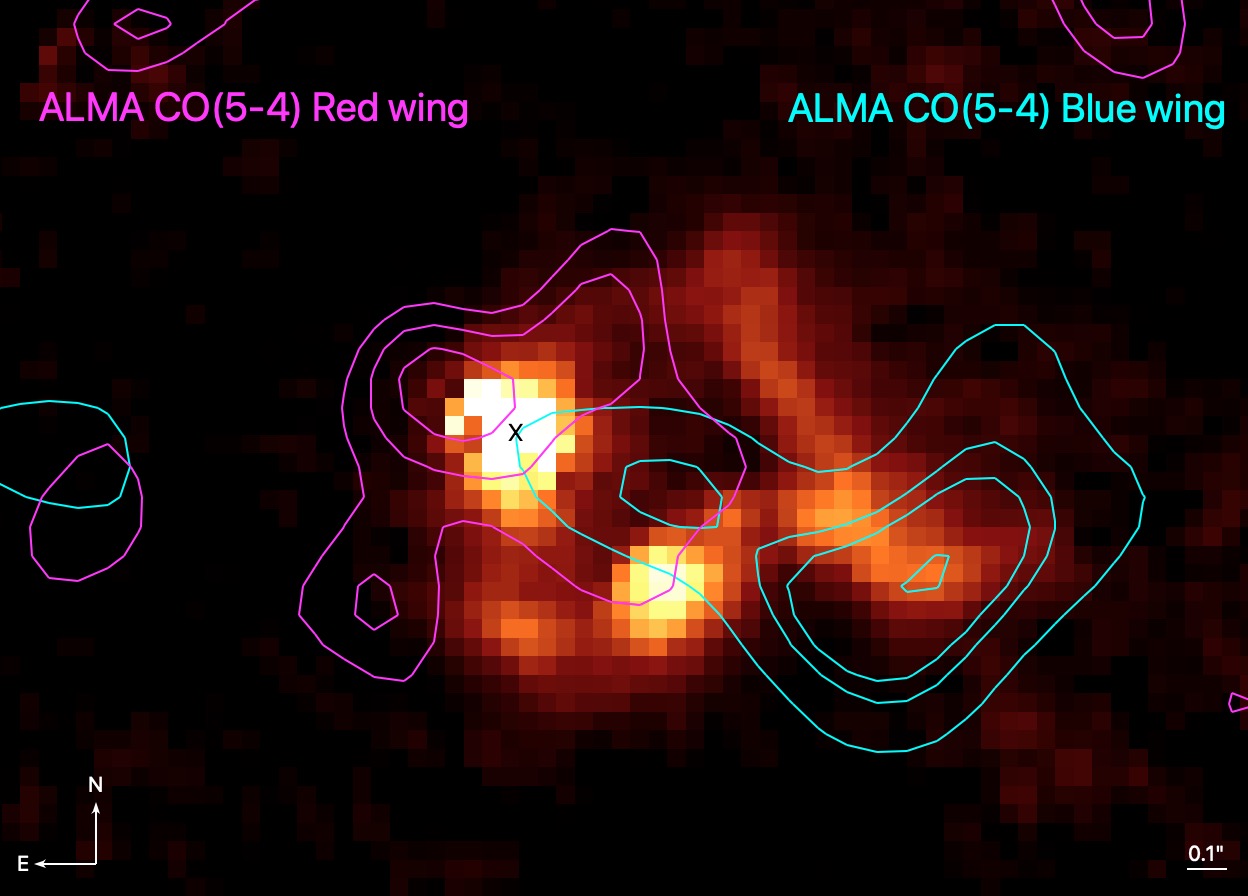}
        \includegraphics[width=0.45\textwidth]{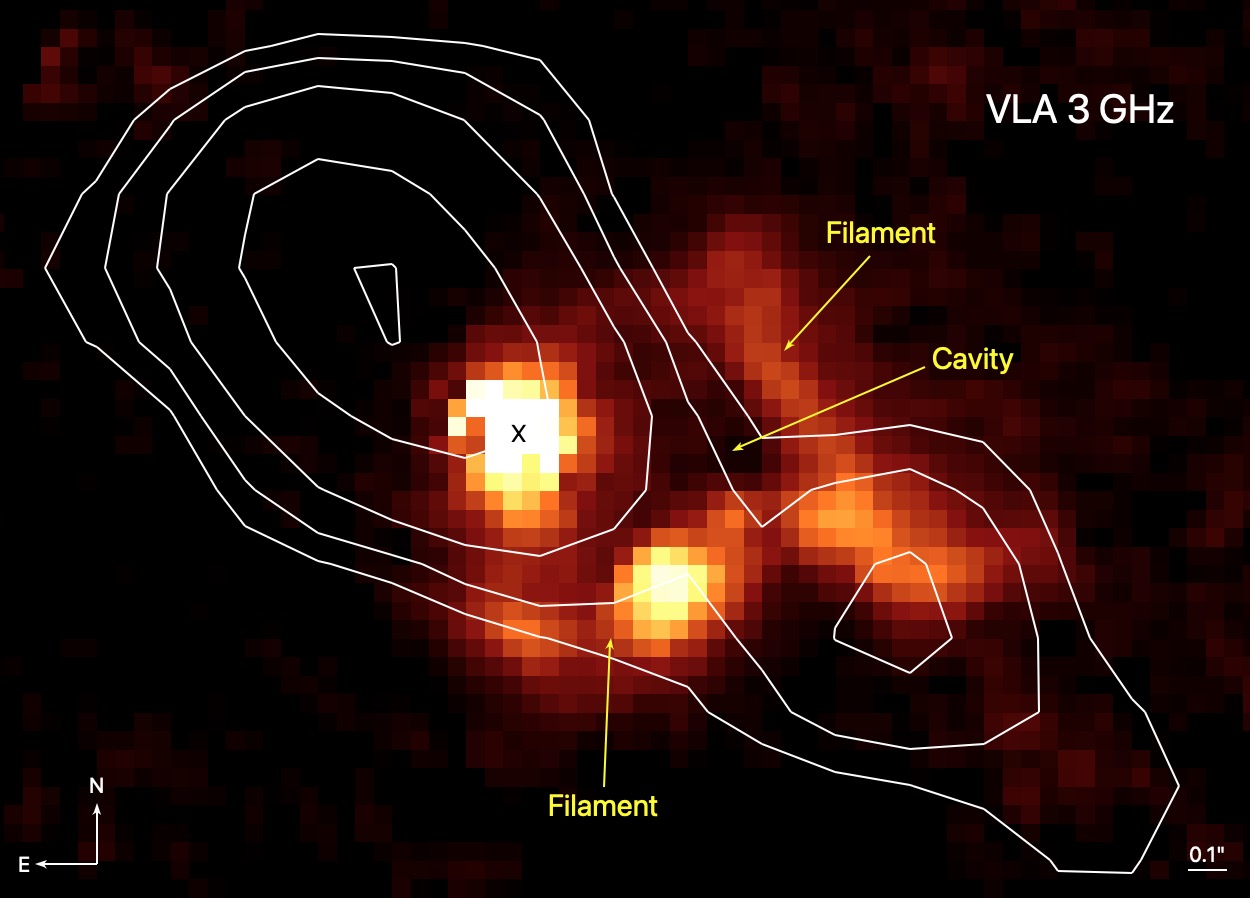}
        \end{center}
        \caption{Channel map of \oiiib\ emission with v$<-300$ km/s with contours of the CO(5-4) molecular blue- (cyan) and redshifted (magenta) outflow components (upper panel, from B18), and  3 GHz radio emission (white) from VLA observations (lower panel, from Vardoulaki et al. \citeyear{vardoulaki19}). The location of the QSO is marked with a black cross, and the filaments and the cavity are shown with arrows. The VLA radio emission and the blueshifted ionised and molecular outflows appear to be co-spatial, as well as the possible redshifted counterparts. 
    } 
        \label{CO_VLA}
\end{figure}
\subsection{Jet-ISM interaction in XID2028} \label{jetism}

Vardoulaki et al. (\citeyear{vardoulaki19}) presented radio images of XID2028 among the multi-components radio sources identified in the VLA-COSMOS Large Project at 3 GHz (0.75" resolution, 2.3 $\mu \textrm{Jy}\ \textrm{beam}^{-1}$ RMS). 
The galaxy is identified as number 10964 in their catalogue, but is surprisingly classified as a star-forming galaxy based on the lack of a radio excess with respect to the IR-radio correlation of star-forming galaxies (Delvecchio et al. \citeyear{delvecchio17}), as well as on the lack of clear jets at 3 GHz and of any compact nuclear radio emission with VLBA observations ($10\ \mu\textrm{Jy}\ \textrm{beam}^{-1}$, at $16.2\times7.3\ \textrm{mas}^2$, Herrera Ruiz et al. \citeyear{herreraruiz17}). In this respect, we note that subsequent deeper VLBA+GBT observations in different parts of the COSMOS field have produced 40\% more new AGN detections (Herrera Ruiz et al. \citeyear{herreraruiz18}), suggesting a possible sensitivity limit of current observations.
\begin{figure*}
        \begin{center}
                \includegraphics[width=0.8\textwidth]{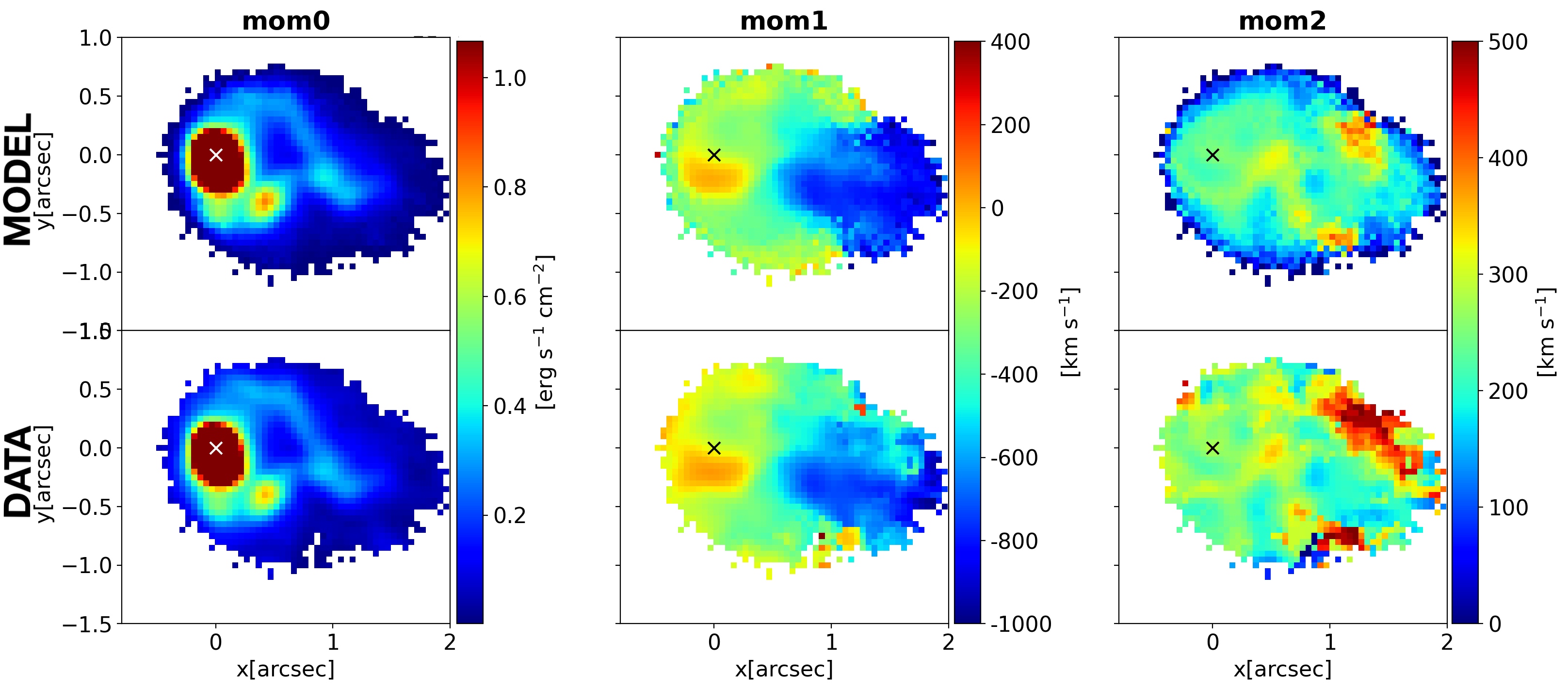}
        \end{center}
        \caption{Comparison between the \MOKA\ weighted model (upper panels) and the NIRSpec [OIII] kinematics (bottom panels) for moment 0, moment 1, and moment 2 in the higher S/N region of the galaxy, showing how the assumed bubble+jet geometry is able to reproduce the observed kinematics. The cross marks the location of the QSO.
    } 
        \label{MOKA3D}
\end{figure*}

However, in the 3 GHz VLA image, the emission peak is not coincident with the QSO and host galaxy centre, as expected if the radio emission were produced by star formation alone (see Fig.~\ref{CO_VLA}, bottom panel). Instead, we observe two blobs that extend north-east and south-west, in opposite directions from the centre, at locations roughly corresponding to the red- and blueshifted outflows detected in the ionised and molecular data (see Fig.~\ref{CO_VLA}, upper panel). The two blobs can thus be interpreted as radio lobes that extend beyond the host galaxy, produced by a bi-polar jet that is co-spatial with the ionised and molecular wind (see e.g. Cresci et al. \citeyear{cresci15b}, Jarvis et al. \citeyear{jarvis19}, Venturi et al. \citeyear{venturi21}). We also note that based on the integrated source flux density at 1.4 GHz (162 $\mu$Jy, Schinnerer et al. \citeyear{schinnerer07}) and at 3 GHz (92 $\mu$Jy, Smol{\v{c}}i{\'c} et al. \citeyear{smolcic17}, Vardoulaki et al. \citeyear{vardoulaki19}), the radio spectral index is $\alpha=0.74$, consistent with synchrotron emission from AGN radio lobes. 
The corresponding radio luminosities, 6.8$\times10^{23}$  W/Hz and 3.9$\times10^{23}$ W/Hz at 1.4 and 3 GHz, respectively, would indeed also be compatible with a low-power radio galaxy. 
%{\bf (no mention to the 610 MHz data point?)}. 
%The corresponding radio luminosities, 6.8$\times10^{23}$  W/Hz and 3.9$\times10^{23}$ W/Hz at 1.4 and 3 GHz, respectively, would be indeed compatible both with low-power FRI radio galaxies and with a radio quiet AGN. Also the latter can be characterized by double-lobe structures interpreted as jet-like outflows (e.g. Leipski et al. \citeyear{leipski06}, Kellermann et al. \citeyear{kellerman16}) 

The coexistence of a bi-polar radio jet with ionised and molecular gas outflows has been observed in several sources (see e.g. Morganti et al. \citeyear{morganti15}, Cresci et al. \citeyear{cresci15b}, Venturi et al. \citeyear{venturi21}, Ulivi et al. in prep., Giroletti et al. \citeyear{giroletti17} and reference therein).
In this scenario, the filamentary structure detected with NIRSpec between the QSO and the outflow can be interpreted as a hot expanding bubble filled with low surface brightness emission that is brightened at the edges, which is inflated and dragged into the host galaxy ISM by the jet. According to the Ks LUCI+ARGOS imaging presented in B18, this is compatible with the host galaxy effective radius $\rm r_e=10$ kpc, as the bubble extends to distances $\rm D_{bubble}=0.7"\sim6$ kpc. Similar structures are also observed in low radio luminosity systems at intermediate redshift (e.g. the Teacup galaxy; Harrison et al. \citeyear{harrison15}, Venturi et al. in prep.). 

Ionised gas filaments surrounding radio lobes have also been observed in higher-luminosity radio galaxies in the nearby Universe, where complex networks of line-emitting streams are observed around the hot X-ray cavities (e.g. 3C317; Balmaverde et al. \citeyear{balmaverde18}). However, in these bright radio galaxies, the filaments show complex kinematics due to the combination of emission from receding and approaching streams. In contrast, the filaments in XID2028 appear to have consistent kinematics, compatible with tracing the brightened edge of an expanding bubble. According to simulations (e.g. Gaibler et al. \citeyear{gaibler12}), the expanding bubble and the jet excavate the central region of the galaxy disk, creating a cavity and possibly triggering star formation  at the edges of the expanding shell by a blast wave causing strong compression and cooling in the ISM (e.g. Silk et al. \citeyear{silk13}, Ishibashi \& Fabian \citeyear{ishibashi12}, C15, Cresci et al. \citeyear{cresci15b}, Cresci \& Maiolino \citeyear{cresci18}, Duggal et al. \citeyear{duggal21}, Bessiere \& Ramos Almeida \citeyear{bessiere22}). %This is consistent with the scenario already proposed by C15, where enhanced emission in \ha\ and in the rest frame U-band associated with star formation was detected in the region corresponding to the filaments at the edge of the expanding bubble. 

At a larger radius ($\sim0.7"$, corresponding to 6 kpc), the jet appears to have pierced the shell, and it propagates faster through the lower-density environment together with the wind. This creates the collimated and highly blueshifted outflows already detected with SINFONI by C15, with approaching projected velocities as high as 1100 km/s. 

A similar scenario is reproduced in 3D hydrodynamic simulations of jets interacting with a clumpy ISM (see e.g. Gaibler et al. \citeyear{gaibler11}, Gaibler et al. \citeyear{gaibler12}, Wagner et al. \citeyear{wagner12}, Dugan et al. \citeyear{dugan17}, Mukherjee et al. \citeyear{mukherjee18}, Hu{\v{s}}ko et al. \citeyear{husko22}). In these simulations, the jet initially inflates energy-driven bubbles in the host galaxy ISM that expand perpendicularly to the galaxy disk. The jet eventually pierces the bubble and propagates into the halo. The asymmetry observed between the approaching and receding side of the outflow can also be ascribed to differences in the ISM density and clumpiness between the two sides of the bipolar jet (Gaibler et al. \citeyear{gaibler11}).  

%
%\begin{figure}
%       \begin{center}
%               \includegraphics[width=0.45\textwidth]{MOKA3D_unweighted.jpg}
%       \end{center}
%       \caption{Upper panel: unweighted Moment 1 map for the best fitting MOKA3D used to reproduce the XID2028 bubble+jet kinematics in Fig.~\ref{MOKA3D}. , and schematic 3D view of the assumed toy model geometry (right panel). The QSO position is shown with a star, while the arrows show the outflow direction together with the expansion velocity inflating the bubble (see text). 
%    } 
%       \label{MOKA3Dunw}
%\end{figure}
%
\begin{figure}
        \begin{center}
                \includegraphics[width=0.45\textwidth]{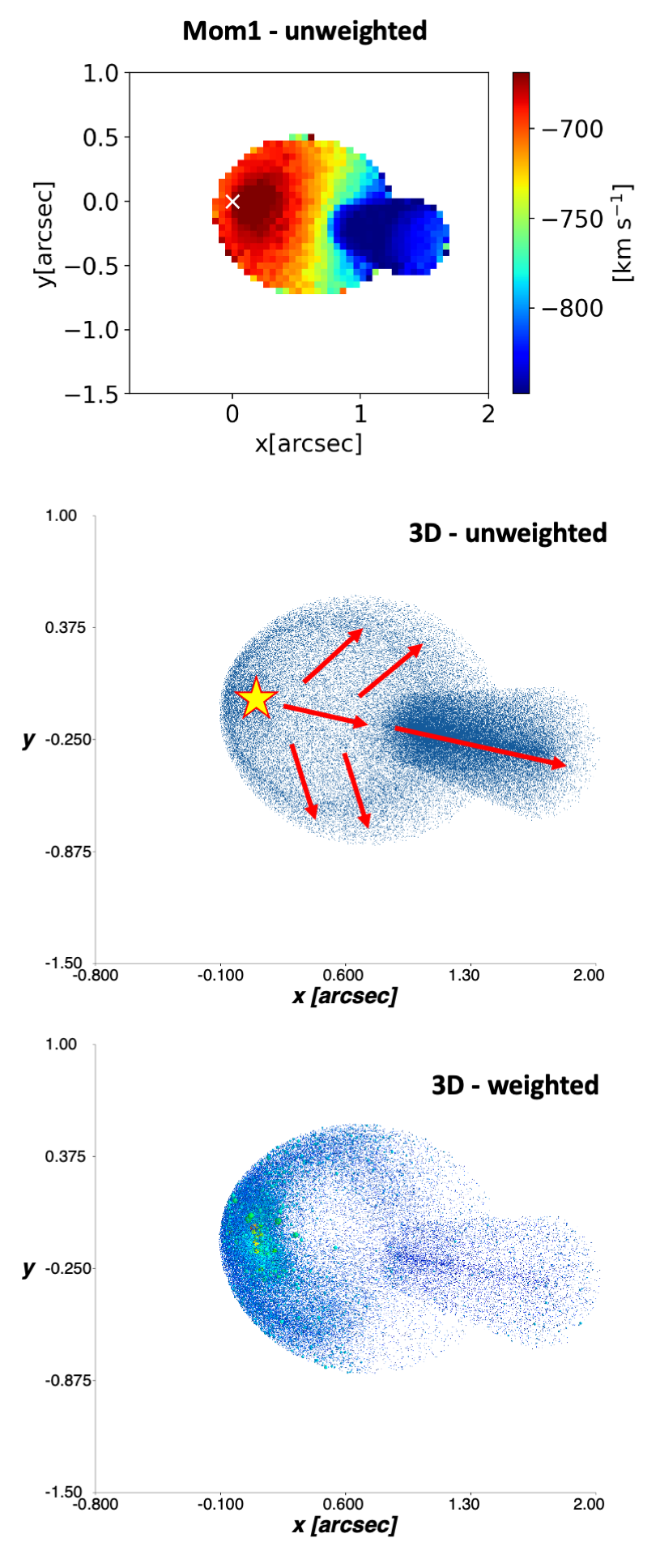}
        \end{center}
        \caption{Best-fitting \MOKA\ model used to reproduce the XID2028 bubble+jet kinematics in Fig.~\ref{MOKA3D}. Upper panel: Projected unweighted moment 1 map. Middle panel: 3D unweighted model, i.e. the intrinsic cloud distribution before weighting them on the observed datacube. A schematic 3D view of the assumed toy-model geometry and kinematics is superposed. The QSO position is shown with a star, and the arrows show the outflow direction together with the expansion velocity that inflates the bubble (see text). Lower panel: 3D model. Each cloud is weighted based on the observed flux for its location and velocity in the data.
    } 
        \label{MOKA3Dunw}
\end{figure}

\subsection{Modelling the observed gas kinematics with \MOKA}

To test whether the shell and piercing jet geometry predicted by simulations can be applied to XID2028, we compared the NIRSpec XID2028 data to the 3D AGN outflow model generated using the modelling framework called modelling outflows kinematics in AGN 3D (\MOKA) that is presented in Marconcini et al. (in prep.). The model assumes a simple velocity field described by an analytical expression, but populates it with a 3D distribution of emitting clouds weighted using the observed line flux in all spaxels and spectral channels of the data cube. This procedure improves upon the assumption of a smooth distribution of gas-emitting regions used in the literature because it is able to reproduce the clumpy appearance observed in the data. Details of the model implementation are discussed in Marconcini et al. (in prep.). Here we just show that the observed gas geometry can be reproduced using an expanding bubble dragged by the jet and the wind, plus a collimated outflow where the radio jet escapes the host galaxy ISM. A comparison between the \MOKA\ model with observations is presented in Fig.~\ref{MOKA3D}, highlighting that this configuration can simultaneously reproduce the filaments, the cavity and the collimated and fast outflow beyond the bubble in flux and velocity. 
The best-fitting unweighted 3D model, that is, the starting model with the intrinsic cloud distribution before weighting the clouds on the observed datacube, is shown in Fig.~\ref{MOKA3Dunw} (middle panel), along with its 2D velocity distribution (upper panel). The weighted model, in which each cloud has been weighted based on the observed flux for its location and velocity in the data, is shown in the bottom panel.
%The  schematic of the assumed bubble+jet geometry, along with the unweighted, smooth velocity distribution for the best estimate solution before weighting each cloud on the observed cube, are shown in Fig.~\ref{MOKA3Dunw}. 

The estimate of the intrinsic deprojected outflow velocity in the best-fitting model is $\textrm{v}_{out,\MOKA}=1027$ km/s, with an aperture angle $\alpha=10^{\circ}$, while the bubble has a radial velocity of $\textrm{v}_{rad,\MOKA}=509$ km/s, an expansion, a tangential velocity $\textrm{v}_{exp,\MOKA}=517$ km/s, and an intrinsic dispersion velocity  $\textrm{v}_{disp,\MOKA}=117$ km/s. The bubble and outflow system is reproduced with an inclination of $\sim27^{\circ}$ with respect to the observer's line of sight.

\subsection{Outflow energetics and mass outflow rate}

\subsubsection{Ionised outflow} \label{ionout}

Through the new NIRSpec data with their higher resolution and sensitivity, we were also able to revise the estimate of mass outflow rate based on SINFONI data by C15. We can directly measure the electron density in the outflow from the \siiab\ line ratio and used the \ha\ flux associated with the outflow to derive the line-emitting ionised gas mass. Following Cresci et al. (\citeyear{cresci17}), for instance,

\begin{equation}
\textrm{M}_{out}=3.2\cdot 10^5 \left(\frac{\textrm{L}_{out}(\textrm{H}\alpha)}{10^{40}\ \textrm{erg}/\textrm{s}}\right)\ \left(\frac{100\ \textrm{cm}^{-3}}{\textrm{n}_{e,out}}\right)\ \textrm{M}_{\sun},
\end{equation}
where $\textrm{L}_{out}(\textrm{H}\alpha)$ is the \ha\ luminosity of the outflow component, and $\textrm{n}_e$ is the electron density measured in the outflow. %As in Sect.~\ref{disk}
To isolate the outflow emission, we excluded the Gaussian components with a velocity shift |v|<300 km/s (with respect to the systemic velocity of the quasar) and $\sigma<300$ km/s (see Fig.~\ref{samplespec}) from each spaxel, exhibiting disk-like kinematics as shown in Fig.~\ref{disk}. In this way, we derived a total outflow \ha\ flux $\textrm{F}_{out}(\textrm{H}\alpha)\sim1.28\pm0.04 \times 10^{-16}$ erg s$^{-1}$ $\textrm{cm}^{-2}$. We corrected this flux for extinction using the \ha/\hb\ line ratio for the broad component, as derived over the outflow region ($\textrm{A}_{\textrm{V}}=1.1\pm0.45$, see Sect.~\ref{ism}), obtaining $\textrm{F}_{out,corr}(\textrm{H}\alpha)=3\pm1 \times 10^{-16}$ erg s$^{-1}$ $\textrm{cm}^{-2}$. Using the electron density for the outflow component measured on the outflow region, $\textrm{n}_{e,out}=360\pm180\ \textrm{cm}^{-3}$, we derived a total outflow mass  $\textrm{M}_{out}=4.5\pm3\times 10^7\ \textrm{M}_{\sun}$. Assuming an outflow velocity $\textrm{v}_{out}=-1100$ km/s, which is the observed maximum $\textrm{v}_{10}$, consistent with the \MOKA\ intrinsic value, and a radius $\textrm{R}_{out}= 1.6"$ ($\sim$ 13.6 kpc), deprojected using the inclination of $\sim27^{\circ}$ from \MOKA\ to 3.5", we obtain an outflow rate (e.g. B18)
\begin{equation}
    \rm \dot M_{out,ion} =  \textrm{v}_{out}\ \frac{\textrm{M}_{out}}{\textrm{R}_{out}} = 6\pm3\ \textrm{M}_{\sun}/ \textrm{yr}.
\end{equation}
The computed value is consistent with the estimate from the \MOKA\ model, from which we derive a median outflow rate in the outflow volume of $\rm \dot M_{out,ion}(\textrm{\MOKA}) \sim 8\ \textrm{M}_{\sun}/ \textrm{yr}$.

The main difference with respect to the analysis presented in C15 ($\rm \dot M_{out,ion}\sim100$) follows from our measured value of n$_e$, which they assumed to be n$_e=100$ cm$^{-3}$, thus pushing the outflow rate towards  higher values, by the deprojection of the outflow radius based on the \MOKA\ modelling, and by the better definition of the outflow component allowed by the NIRSpec data.
The corresponding kinetic power of the ionised outflow is given by
\begin{equation}
    \rm \dot E_{kin,ion}=\frac{1}{2} \rm \dot M_{out,ion} \textrm{v}^2=2.1\pm1.4\times 10^{42}\ \textrm{erg}\ \textrm{s}^{-1},
\end{equation}
and the momentum rate $\rm \dot P_{kin,ion}=\dot M_{out,ion}\ \textrm{v}_{out}=4\pm2\times 10^{34}$ dyne. 

An estimate of the ionised outflow rate can be also derived from the observed [OIII] flux in the broad component (e.g. Cano-Diaz et al. \citeyear{cano12}).  We measure a dust corrected total [OIII] emission from the outflow $\textrm{F}_{out}(\textrm{[OIII]})\sim10\pm5 \times 10^{-16}\ \textrm{erg}\ \textrm{s}^{-1}\ \textrm{cm}^{-2}$, which translates into a total outflowing mass $\rm M_{out}([OIII])=1.2\pm0.8 \times 10^7\ \textrm{M}_{\sun}$ assuming solar metallicity and the same values for the density as were used for \ha. This translates into $\rm \dot M_{out,ion}([OIII]) = 3\pm2\ \textrm{M}_{\sun}/ \textrm{yr}$, consistent with the \ha\ value, but slightly lower. However, given the more uncertain set of assumptions needed to compute the ionised gas mass from [OIII] (e.g. gas metallicity and ionisation structure; see Carniani et al. \citeyear{carniani15}), we chose to use the \ha\ based measurement as the fiducial mass.  

\subsubsection{Neutral outflow}

\begin{figure}
        \begin{center}
                \includegraphics[width=0.45\textwidth]{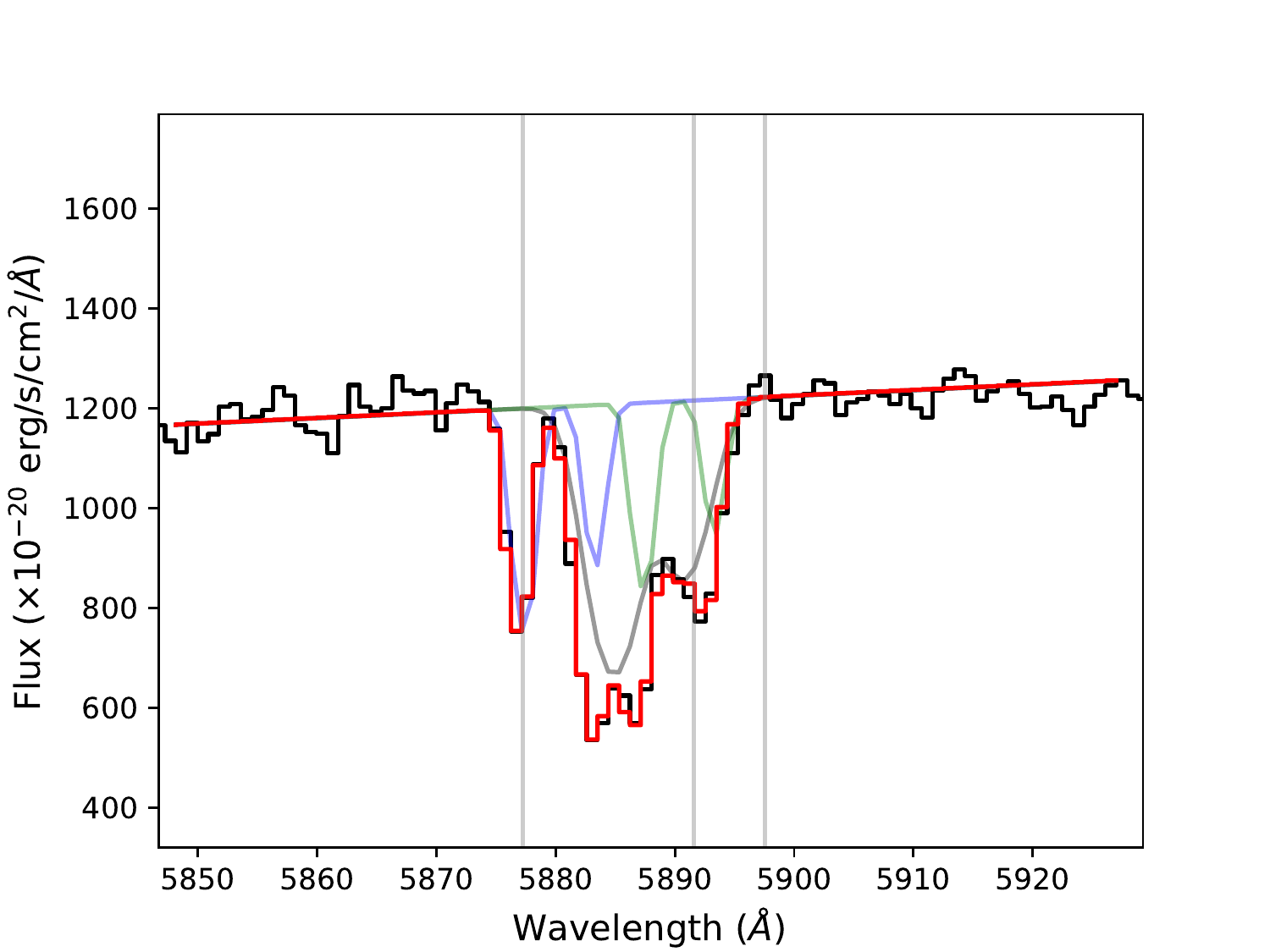}
        \end{center}
        \caption{Portion of NIRSpec integrated spectra of XID2028 including the NaID doublet. The best-fitting components are superposed. The feature has been reproduced with three blueshifted velocity components (see text). The vertical lines indicate the systemic wavelengths of the HeI$\lambda$5877 and NaID$\lambda$5890,5896 doublet lines. 
    } 
        \label{NaIDfit}
\end{figure}

The neutral sodium absorption lines NaID$\lambda\lambda$5890,5896 are clearly detected in the NIRSpec nuclear spectrum. We fitted the observed intensity profiles of the sodium lines in an integrated spectrum on a region centred on the QSO with a radius of 10 spaxels (0.5"), where the absorption is detected,  considering a model parametrised in the optical depth space (e.g. Rupke et al. \citeyear{rupke02}). As for the emission lines, a multiple-component fit was required to reproduce the observed profiles in absorption. The best-fit model reported in Fig. \ref{NaIDfit} shows three distinct blueshifted kinematic components in absorption, at $\rm v_{0,1}=-215$ km/s (FWHM=93 km/s, $\rm log(N_{H,1}/\textrm{cm}^{-2})=20.68$), $\rm v_{0,2}=-260$ km/s (FWHM=234 km/s, $\rm log(N_{H,2}/\textrm{cm}^{-2})=20.89$), and $\rm v_{0,3}=-640$ km/s (FWHM=94 km/s, $\rm log(N_{H,3}/\textrm{cm}^{-2})=20.68$); no emission line components are required for either the NaID or the HeI$\lambda$5877  line to reproduce the observed profile in the fit. Following the prescriptions indicated by Perna et al. (\citeyear{perna15}), we derived a new estimate for the neutral outflow energetics, assuming that the neutral gas flow extends up to 1 kpc, 
\begin{equation}
\rm \dot M_{out,neut} = 7\cdot \sum_{i=1}^3 \left(\frac{\textrm{N}_{\textrm{H},i}}{10^{20}\ \textrm{cm}^{-2}} \right) \left(\frac{\rm R}{5\ \textrm{kpc}}\right) \left(\frac{\textrm{v}_{0,i}}{300\ \textrm{km}/\textrm{s}}\right) \sim 30\ \textrm{M}_\odot/\textrm{yr},
\end{equation}
where we sum the contributions of the three different velocity components.  
This measurement should be considered as an order-of-magnitude estimate because we cannot constrain the spatial extension of the neutral outflow on the line of sight; nevertheless, this mass rate is broadly consistent with the lower limit reported in  Perna et al. (\citeyear{perna15}), $\rm \dot M_{out,neut}\gtrsim 80\ \textrm{M}_{\sun}/\textrm{yr}$, based on MgII absorption in their X-Shooter spectra after accounting for the different assumed outflow radius (in Perna et al. \citeyear{perna15} the same extension as the ionised outflow, 11 kpc).

\subsubsection{Total outflow rate}

The molecular outflow phase has been studied in detail by B18. Rescaling their outflow rate taking the increased deprojected radius based on the \MOKA\ inclination into account, we derived a molecular outflow rate $\rm \dot M_{out,mol}\sim20-120\ \textrm{M}_{\sun}/\textrm{yr}$ depending on the assumed CO-H$_2$ conversion factor $\alpha_{\rm CO}$ (see B18 for details). %, while Perna et al. (\citeyear{perna15}) found a lower limit for the neutral gas outflow $\dot M_{out,neut}\gtrsim 80\ \textrm{M}_{\sun}/\textrm{yr}$ based on MgII absorption in their X-Shooter spectra. 
Summing up all the three gas phases, we derive a total mass outflow rate
\begin{equation}
    \rm \dot M_{out,tot}\sim60-160\ \textrm{M}_{\sun}/\textrm{yr}.
\end{equation}

Based on this revised value for the total outflow rate and using the molecular gas mass derived by B18 ($\textrm{M}_{gas}=1.1\pm0.5 \times 10^{10}\ \textrm{M}_{\sun}$) as well as the SFR$\sim250\ \textrm{M}_{\sun}/\textrm{yr}$ derived from the total IR luminosity (Perna et al. \citeyear{perna15}), we can estimate the total depletion time for this source. Using the numbers above, we derive $\rm \tau_{depl}=\textrm{M}_{gas}/(\textrm{SFR}+\dot M_{out,tot})\sim30$ Myr. This timescale is remarkably shorter than typical gas depletion times in normal star-forming galaxies with comparable stellar mass and redshift ($\sim0.5-1.5$ Gyr; see e.g. Sargent et al. \citeyear{sargent14}, Tacconi et al. \citeyear{tacconi18}). This further supports the significant effect of AGN feedback in this source. 

\section{Conclusions} \label{conclusions}

We have presented JWST NIRSpec IFU observations of XID2028, a prototypical outflowing obscured QSO at z=1.59.  A massive and extended outflow has previously been detected in this source in different gas phases in previous SINFONI (ionised), X-Shooter (neutral atomic), and ALMA (neutral molecular) observations. The unprecedented sensitivity and resolution of NIRSpec allowed us to reveal new details of the morphology, kinematics, and wind/jet interaction with the ISM in this iconic source. In particular:
\begin{itemize}
    \item we detected a rotational pattern in the narrowest component of the emission lines, which we interpreted as tracing a rotating disk in the host galaxy (see Fig.~\ref{rotation}). The velocity gradient is consistent with the inner kinematics observed in CO(5-4) line emission observed with ALMA (B18), supporting the estimate of a dynamical mass $\textrm{M}_{\textrm{dyn}}\sim10^{11}$ M$_{\sun}$ within the observed CO and ionised gas rotating structure.
    \item A highly blueshifted and collimated outflow, with velocities up to $\sim-1100$ km/s, is detected  $\sim6$ kpc apart from the QSO, extending up to $\sim13$ kpc in projection (see the upper left panel of Fig.~\ref{channels}). The ionised wind is co-spatial with the molecular outflow detected in B18 and consistent with the detection in [OIII] by C15. A possible redshifted outflow component is also detected in the ionised phase. It is spatially consistent with the molecular redshifted outflow counterpart in the opposite direction with respect to the QSO and the blueshifted outflow, with velocities up to $\sim 700$ km/s (see the lower right panel of Fig.~\ref{channels} and the upper panel of Fig.~\ref{CO_VLA}).
    \item The outflow appears to be connected to the QSO by a system of two line-emitting filaments, surrounding a region with suppressed  emission (see the moment 0 map in Fig.~\ref{gaskin}), resembling shell-like features that have been observed in local low-luminosity radio galaxies. The filaments have low projected velocities close to the QSO, and show progressively more blueshifted velocity in regions closer to the fast collimated outflow (see the moment 1 map in Fig.~\ref{gaskin}).
    \item Archival VLA 3 GHz observations %(Vardoulaki et al. \citeyear{vardoulaki19}) 
    show two extended radio lobes that are broadly coincident with the bi-polar outflows, suggesting low-luminosity radio jets in the galaxy (see Fig.~\ref{CO_VLA}). We thus interpret the filamentary structure between the QSO and the outflow as a hot expanding bubble filled with low surface brightness emission brightened at the edges that is inflated and dragged into the host galaxy ISM by the jet. At a larger distance from the QSO ($\sim6$ kpc projected), the jet pierces the shell, and it propagates faster through the lower-density environment.
    \item We compared the NIRSpec data with the expectation from our \MOKA\ (Marconcini et al. in prep.) AGN outflow 3D model assuming this expanding bubble + collimated outflow geometry. We showed that this scenario can reproduce the observed kinematics (see Fig.~\ref{MOKA3D}).
    \item The proposed scenario is compatible with a contribution of both negative feedback along the path of the outflow, where the gas is heated and removed from the host galaxy, and positive feedback at the bubble edges, where the compression of the ISM during the expansion can trigger star formation within the feedback-driven shell, as traced by rest-frame U-band emission. However, we do not detect a residual \ha\ component ascribable to star formation in the host.
    \item We recomputed the ionised outflow energetics and mass outflow rate by directly measuring the extinction and electron density in the outflow (presented in Fig.~\ref{avne}) and based on the more detailed kinematics and structural information allowed by the new NIRSpec data. We derived an ionised gas outflow rate $\rm \dot M_{out,ion}=6\pm3\ \textrm{M}_{\sun}/\textrm{yr}$ and a kinetic power $\rm \dot E_{kin,ion}\sim 2\times10^{42}\ erg\ s^{-1}$.
    \item We estimated a neutral outflow rate of $\rm \dot M_{out,neut} \sim 30\ \textrm{M}_\odot/\textrm{yr}$ based on the detection of blueshifted velocity components of the NaID doublet on the nuclear region (see Fig.~\ref{NaIDfit}). Although the estimate is uncertain due to the unknown extent of the absorbing material, the results confirm the estimate by Perna et al. (\citeyear{perna15}) based on MgII absorption.
    \item Including the contribution of the molecular outflow from B18 in the ionised and neutral phases derived by NIRSpec, we derived a total mass outflow rate of $\rm \dot M_{out,tot}\sim110\ \textrm{M}_{\sun}/\textrm{yr}$. This corresponds to a depletion time of just $\rm \tau_{depl}=\textrm{M}_{gas}/(\textrm{SFR}+\dot M_{out,tot})\sim30$ Myr, which supports the key role of AGN feedback in this source.
\end{itemize}

These observations add to the growing evidence that the radio emission and ionised gas outflows in active galaxies are connected. The also show the relevant role of radio jets in AGN feedback on the host galaxy ISM even at modest luminosities (see e.g. Jarvis et al. \citeyear{jarvis19}, \citeyear{jarvis21}, Venturi et al. \citeyear{venturi21}, Girdhar et al. \citeyear{girdhar22}). 

Altogether, the NIRSpec observations analysed in this work demonstrate that NIRSpec is capable of providing unprecedented information on the role of quasar feedback at high redshift. Future NIRSpec and MIRI datasets will be thus fundamental not only to enable our first glance at AGN feedback at z$>$4, but also to overcome the current limitations of ground-based IFU in the study of QSOs at z$\sim1.5-2.5$. They will provide the required major leap in resolution and sensitivity to obtain a sharper answer regarding the role of quasar feedback on galaxy evolution at the epoch of peak of AGN activity in the Universe, the golden epoch for AGN feedback.

%--------------------------------------------------------------------

\begin{acknowledgements}
GC, GT, MP, MB, AM, FM, FB and GV acknowledge the support of the INAF Large Grant 2022 "The metal circle: a new sharp view of the baryon cycle up to Cosmic Dawn with the latest generation IFU facilities". 
GC, MB, AM and MB acknowledge support from PRIN MIUR project “Black Hole winds and
the Baryon Life Cycle of Galaxies: the stone-guest at the galaxy evolution supper”, contract \#2017PH3WAT. 
FM, AM, and GC acknowledge support from the INAF Large Grant 2022 "Dual and binary SMBH in the multi-messenger era". 
SA and MP acknowledge support from the research project PID2021-127718NB-I00 of the Spanish Ministry of Science and Innovation/State Agency of Research (MICIN/AEI).
%MB acknowledges support from the agreement ASI-INAF n. 2017-14-H.O. 
GV acknowledges support from ANID program FONDECYT Postdoctorado 3200802.
H{\"U} gratefully acknowledges support by the Isaac Newton Trust and by the Kavli Foundation through a Newton-Kavli Junior Fellowship. 
Funded by the European Union (ERC, WINGS, 101040227). Views and opinions expressed are however those of the author(s) only and do not necessarily reflect those of the European Union or the European Research Council Executive Agency. Neither the European Union nor the granting authority can be held responsible for them.
\end{acknowledgements}

% WARNING
%-------------------------------------------------------------------
% Please note that we have included the references to the file aa.dem in
% order to compile it, but we ask you to:
%
% - use BibTeX with the regular commands:
%   \bibliographystyle{aa} % style aa.bst
%   \bibliography{Yourfile} % your references Yourfile.bib
%
% - join the .bib files when you upload your source files
%-------------------------------------------------------------------

\end{document}